\documentclass[10pt,twoside,a4paper]{article}
\makeatletter\if@twocolumn\PassOptionsToPackage{switch}{lineno}\else\fi\makeatother

      \makeatletter
\usepackage{wrapfig}
\newcounter{aubio}

\long\def\bioItem{%
\@ifnextchar[{\@bioItem}{\@@bioItem}}

\long\def\@bioItem[#1]#2#3{
 \stepcounter{aubio}
 \expandafter\gdef\csname authorImage\theaubio\endcsname{#1}
 \expandafter\gdef\csname authorName\theaubio\endcsname{#2}
 \expandafter\gdef\csname authorDetails\theaubio\endcsname{#3}
}

\long\def\@@bioItem#1#2{
 \stepcounter{aubio}
 \expandafter\gdef\csname authorName\theaubio\endcsname{#1}
 \expandafter\gdef\csname authorDetails\theaubio\endcsname{#2}
}

\newcommand{\checkheight}[1]{%
  \par \penalty-100\begingroup%
  \setbox8=\hbox{#1}%
  \setlength{\dimen@}{\ht8}%
  \dimen@ii\pagegoal \advance\dimen@ii-\pagetotal
  \ifdim \dimen@>\dimen@ii
    \break
  \fi\endgroup}

\def\printBio{%
  \@tempcnta=0
   \loop
     \advance \@tempcnta by 1
     \def\aubioCnt{\the\@tempcnta}
     \setlength{\intextsep}{0pt}%
     \setlength{\columnsep}{10pt}%
     \expandafter\ifx\csname authorImage\aubioCnt\endcsname\relax%
      \else%
       \checkheight{\includegraphics[height=1.25in,width=1in,keepaspectratio]{\csname authorImage\aubioCnt\endcsname}}
        \begin{wrapfigure}{l}{25mm}
         \includegraphics[height=1.25in,width=1in,keepaspectratio]{\csname authorImage\aubioCnt\endcsname}
        \end{wrapfigure}\par
      \fi
     \noindent\textbf{\csname authorName\aubioCnt\endcsname}\csname authorDetails\aubioCnt\endcsname \par\bigskip
      \ifnum\@tempcnta < \theaubio
   \repeat
   }
\makeatother

\usepackage{amsfonts,amssymb,amsbsy,latexsym,amsmath,tabulary,graphicx,times,xcolor}

\usepackage[T1]{fontenc}

\usepackage{caption,fancyhdr,abstract,textcase}
\usepackage[paperheight=297mm,paperwidth=210mm,margin=30mm,left=30mm,right=30mm,top=30mm]{geometry}
\usepackage{textcase}
\makeatletter
\def\oupIndent{1pt}
\def\author#1{\gdef\@author{\hskip-\dimexpr(\tabcolsep)\hskip1pt\parbox{\dimexpr\textwidth-1pt}{\centering{\fontsize{13pt}{15.6pt}\selectfont  #1}}}}

\def\title#1{\gdef\@title{\vspace*{-3pc}\bfseries\centering\ifx\@articleType\@empty\else\@articleType\\\fi {\fontfamily{ppl}\fontsize{20pt}{24pt}\selectfont\MakeTextUppercase{ #1} \vspace*{24pt}}}}

\fancypagestyle{headings}{\fancyhf{}\fancyfoot[R]{}}

\fancypagestyle{plain}{\fancyhf{}\fancyfoot[R]{}}

\let\@articleType\@empty \def\articletype#1{\gdef\@articleType{{\normalfont\underline{#1}}}}

\def\abstractname{\textbf{\textit{A\small{BSTRACT}}}}

\renewenvironment{abstract} {\trivlist\item[]\leftskip\oupIndent\par\vskip4pt\noindent{\fontsize{13pt}{15.6pt}\selectfont\textit{\scshape\abstractname}}\mbox{\null}\vspace{5pt}\\ \itshape\fontsize{10pt}{12pt}\selectfont}{\noindent\endtrivlist}

\usepackage[explicit]{titlesec}
\def\NormalBaseline{\def\baselinestretch{1.1}}
\titleformat{\section}[hang]{\NormalBaseline\filright\large\scshape\bfseries\fontsize{14}{16.8}\selectfont}
{\fontsize{14}{16.8}\selectfont\thesection.}
{2pt}
{#1}
[]
\titleformat{\subsection}[hang]{\NormalBaseline\filright\bfseries\fontsize{12}{14.4}\selectfont}
{\thesubsection.}
{2pt}
{#1}
[]
\titleformat{\subsubsection}[hang]{\NormalBaseline\filright\bfseries\fontsize{11}{13.2}\selectfont}
{\thesubsubsection.}
{2pt}
{#1}
[]
\titleformat{\paragraph}[runin]{\NormalBaseline\filright\itshape\fontsize{11}{13.2}\selectfont}
{\theparagraph}
{2pt}
{#1}
[\unskip.]
\titleformat{\subparagraph}[runin]{\NormalBaseline\filright\fontsize{11}{13.2}\selectfont}
{\thesubparagraph}
{2pt}
{#1}
[\unskip.]

\titlespacing{\section}{0pt}{1.5\baselineskip}{.2\baselineskip}
\titlespacing{\subsection}{0pt}{1.5\baselineskip}{.2\baselineskip}
\titlespacing{\subsubsection}{0pt}{1.5\baselineskip}{.2\baselineskip}
\titlespacing{\paragraph}{0pt}{.5\baselineskip}{10pt}
\titlespacing{\subparagraph}{0pt}{.5\baselineskip}{10pt}

\makeatother

\date{}

\usepackage{url,multirow,morefloats,floatflt,cancel,tfrupee}
\makeatletter

\AtBeginDocument{\@ifpackageloaded{textcomp}{}{\usepackage{textcomp}}}
\makeatother
\usepackage{colortbl}
\usepackage{xcolor}
\usepackage{pifont}
\usepackage[nointegrals]{wasysym}
\makeatletter

\def\mcWidth#1{\csname TY@F#1\endcsname+\tabcolsep}

\def\cAlignHack{\rightskip\@flushglue\leftskip\@flushglue\parindent\z@\parfillskip\z@skip}
\def\rAlignHack{\rightskip\z@skip\leftskip\@flushglue \parindent\z@\parfillskip\z@skip}

\usepackage{ifxetex}
\ifxetex\else\if@twocolumn\usepackage{dblfloatfix}\fi\fi

\AtBeginDocument{
\expandafter\ifx\csname eqalign\endcsname\relax
\def\eqalign#1{\null\vcenter{\def\\{\cr}\openup\jot\m@th
  \ialign{\strut$\displaystyle{##}$\hfil&$\displaystyle{{}##}$\hfil
      \crcr#1\crcr}}\,}
\fi
}

\AtBeginDocument{%
  \@ifpackageloaded{endfloat}%
   {\renewcommand\efloat@iwrite[1]{\immediate\expandafter\protected@write\csname efloat@post#1\endcsname{}}}{\newif\ifefloat@tables}%
}%

\def\BreakURLText#1{\@tfor\brk@tempa:=#1\do{\brk@tempa\hskip0pt}}
\let\lt=<
\let\gt=>
\def\processVert{\ifmmode|\else\textbar\fi}

\@ifundefined{subparagraph}{
\def\subparagraph{\@startsection{paragraph}{5}{2\parindent}{0ex plus 0.1ex minus 0.1ex}%
{0ex}{\normalfont\small\itshape}}%
}{}

\newcommand\role[1]{\unskip}
\newcommand\aucollab[1]{\unskip}

\@ifundefined{tsGraphicsScaleX}{\gdef\tsGraphicsScaleX{1}}{}
\@ifundefined{tsGraphicsScaleY}{\gdef\tsGraphicsScaleY{.9}}{}
\def\checkGraphicsWidth{\ifdim\Gin@nat@width>\linewidth
	\tsGraphicsScaleX\linewidth\else\Gin@nat@width\fi}

\def\checkGraphicsHeight{\ifdim\Gin@nat@height>.9\textheight
	\tsGraphicsScaleY\textheight\else\Gin@nat@height\fi}

\def\fixFloatSize#1{}
\let\ts@includegraphics\includegraphics

\def\inlinegraphic[#1]#2{{\edef\@tempa{#1}\edef\baseline@shift{\ifx\@tempa\@empty0\else#1\fi}\edef\tempZ{\the\numexpr(\numexpr(\baseline@shift*\f@size/100))}\protect\raisebox{\tempZ pt}{\ts@includegraphics{#2}}}}

\AtBeginDocument{\def\includegraphics{\@ifnextchar[{\ts@includegraphics}{\ts@includegraphics[width=\checkGraphicsWidth,height=\checkGraphicsHeight,keepaspectratio]}}}

\DeclareMathAlphabet{\mathpzc}{OT1}{pzc}{m}{it}

\def\URL#1#2{\@ifundefined{href}{#2}{\href{#1}{#2}}}

\def\UrlOrds{\do\*\do\-\do\~\do\'\do\"\do\-}%
\g@addto@macro{\UrlBreaks}{\UrlOrds}

\@ifundefined{quoteAttrib}
	{}
	{}

\@ifundefined{titlequoteAttrib}
	{}{}

\newenvironment{title-quote}
	{\list{}{\fontsize{10pt}{12pt}\selectfont\leftmargin.5in\itshape\rightmargin\leftmargin}%
  \item\relax}
  {\endlist}

\makeatother

\usepackage{supertabular}
\usepackage{graphicx}

\usepackage{longtable}
\usepackage{amsmath}
\usepackage{tabularx}

\usepackage{grffile}  

\usepackage{caption}
\usepackage{subcaption}

\usepackage{chronology}
\usepackage{algorithmic}
\usepackage[]{algorithm2e}
\usepackage{amssymb}
\usepackage{float}
\usepackage{threeparttable}
\usepackage[numbers,sort&compress]{natbib}

\usepackage{array}
\usepackage{subcaption}
\usepackage{multirow}
\usepackage{rotating}
\usepackage{pdflscape}

\usepackage{amssymb}

\begin{document}

\title{\huge{SLA-Driven Load Scheduling in Multi-Tier Cloud Computing: Financial Impact Considerations}}
\author{\Large{Husam Suleiman~and~Otman Basir}\\
\large{Department of Electrical and Computer Engineering, University of Waterloo}\\
\large{husam.suleiman@uwaterloo.ca,~obasir@uwaterloo.ca}
}

\def\journalTitle{International Journal on Cloud Computing: Services and Architecture (IJCCSA)}

\maketitle

\begin{abstract}

A cloud service provider strives to effectively provide a high Quality of Service (QoS) to client jobs. Such jobs vary in computational and Service-Level-Agreement (SLA) obligations, as well as differ with respect to tolerating delays and SLA violations. The job scheduling plays a critical role in servicing cloud demands by allocating appropriate resources to execute client jobs. The response to such jobs is optimized by the cloud service provider on a multi-tier cloud computing environment.
Typically, the complex and dynamic nature of multi-tier environments incurs difficulties in meeting such demands, because tiers are dependent on each others which in turn makes bottlenecks of a tier shift to escalate in subsequent tiers.
However, the optimization process of existing approaches produces single-tier-driven schedules that do not employ the differential impact of SLA violations in executing client jobs. Furthermore, the impact of schedules optimized at the tier level on the performance of schedules formulated in subsequent tiers tends to be ignored, resulting in a less than optimal performance when measured at the multi-tier level. Thus, failing in committing job obligations incurs SLA penalties that often take the form of either financial compensations, or losing future interests and motivations of unsatisfied clients in the service provided.
Therefore, tolerating the risk of such delays on the operational performance of a cloud service provider is vital to meet SLA expectations and mitigate their associated commercial penalties. Such situations demand the cloud service provider to employ scalable service mechanisms that efficiently manage the execution of resource loads in accordance to their financial influence on the system performance, so as to ensure system reliability and cost reduction.
In this paper, a scheduling and allocation approach is proposed to formulate schedules that account for differential impacts of SLA violation penalties and, thus, produce schedules that are optimal in financial performance. A queue virtualization scheme is designed to facilitate the formulation of optimal schedules at the tier and multi-tier levels of the cloud environment. Because the scheduling problem is NP-hard, a biologically inspired approach is proposed to mitigate the complexity of finding optimal schedules. The reported results in this paper demonstrate the efficacy of the proposed approach in formulating cost-optimal schedules that reduce SLA penalties of jobs at various architectural granularities of the multi-tier cloud environment.

\end{abstract}

\section{Introduction}
\label{sec:intro}

In a cloud computing environment, client jobs have different service demands and QoS obligations that should be met by the cloud service provider. The arrival of such jobs tends to be random in nature. Cloud resources should deliver services to fulfill different client demands, yet such resources might be limited. Arrival rates of jobs dynamically vary at run-time, which in turn cause bottlenecks and execution difficulties on cloud resources. It is typical that an SLA is employed to govern the QoS obligations of the cloud computing service provider to the client. A service provider conundrum revolves around the desire to maintain a balance between two conflicting objectives: the limited resources available
for computing and the high QoS expectations of varying random computing demands. Any imbalance in managing these conflicting objectives may result in either dissatisfied clients and potentially significant commercial penalties, or an over-sourced cloud computing environment with large assets of computational resources that can be significantly costly to acquire and operate.

Various scheduling approaches are presented in the literature to address the problem so that QoS expectations of client jobs are obtained. Such approaches often focus on optimizing system-level metrics at the resource level of the cloud computing environment, and hence aim at minimizing the response times of client jobs by allocating adequate resources. The response time of a job entails two components: the job's waiting time at the queue level and the job's service time at the resource level. The bottleneck of jobs in the queues has a direct impact on the waiting times of client jobs and, thus, their response times.

A major limitation in schedulers of existing approaches is that they often optimize performance of schedules at the individual resource level. As such, they fail to take advantage of any available capacities of the other resources within the tier. Furthermore, single-resource-driven scheduling is blind to the impact of the resultant schedules on other tiers. Due to complications of the bottleneck shifting and dependencies between tiers of the multi-tier cloud environment, SLA violations of client jobs in a tier would escalate when such jobs progress through subsequent tiers of the cloud environment. Also, such schedules are blind to penalties incurred by the cloud service provider due to SLA violations.

It is typical that a cloud service provider strives to maintain the highest QoS provided to clients, so as to maintain client satisfaction~\cite{Shawish2014, Chana2014QoSSLA, Vuong2015disasterM}. The more satisfied the clients, the higher the likelihood they will choose the cloud service provider to execute their demands. However, cloud jobs often differ with respect to delay tolerance, resource computational demand, QoS expectations, and financial value. Furthermore, certain jobs are time-critical and hence cannot tolerate execution delays, as well as are financially delay-sensitive and tightly coupled with the client experience. Any delays in responding to SLA obligations of such jobs would likely cause financial losses and negative reputation consequences, which thus negatively affects client loyalties of choosing the cloud service provider.

Take, for example, the first notice of a loss application. Once a vehicle gets into an accident, an on-board system detects and sends the accident data to the cloud service provider to process and determine accident location severity, and as a result, notify the appropriate police department. Any delay in processing these data leads to catastrophic consequences. Thus, the SLA that governs this application produces severe penalties reflective of these consequences.

Therefore, the cloud service provider must: (1) ensure resource availability for such jobs under all circumstances, which has to be a function of SLA impacts associated with the jobs; (2) formulate cost-optimal schedules that account for the differential impact of delays in executing client jobs to minimize potential penalties due to such delays; (3) develop a model that computes SLA violation penalties of client jobs and supports the commitment of the cloud service provider in delivering better service and client experience; (4) mitigate the computational complexity of scheduling the excessive client demands on resource queues, as well as facilitate the exploration and exploitation through the search space of schedules to find an optimal scheduling solution.

In this paper, a differentiated impact scheduling approach is proposed to formulate penalty-aware QoS-driven schedules that are optimal in financial performance. The scheduling approach extends the previous work published in~\cite{SuleimanP1_2019, SuleimanP2_2019}, and accounts for the followings:
\begin{itemize}
  \item The utilization of resources within a tier is leveraged so as to influence tier-driven schedules that account for the mutual performance impact of tier resources on the system performance.
  \item The effect of tier dependencies on the system performance is leveraged so as to produce multi-tier-driven schedules that contemplate the impact of schedules optimized in a tier on the performance of schedules formulated in subsequent tiers.
  \item A penalty model that allows for differential treatments of jobs is employed so as to ensure financially optimal job schedules.
  \item A genetic-based approach and a queue virtualization scheme are designed to formulate schedules at the tier and multi-tier levels of the cloud environment, as well as to alleviate and simplify the complexity of finding optimal schedules.
\end{itemize}

\section{Background and Related Work}
\label{sec:backRelated}

The performance of a cloud service provider is highly influenced by the availability of resources, to ensure reliable and efficient executions of the varying client demands. Improving the cloud performance through efficient service models is a driving factor to properly tackle the differentiated QoS penalties of client jobs, so as to reduce negative commercial consequences on the cloud service provider and the client~\cite{SLAclause, cochran2011governance, SLAriskManage2012}. Such models should maintain high client satisfactions and business continuity through reliable scheduling and balancing strategies that support efficient utilization of cloud resources. The strategies should provide services to client demands in a timely manner and thus mitigate the negative impact on the QoS delivered to clients~\cite{Quality2018}.

Existing approaches in the literature typically address the scheduling of client jobs that entail identical SLA penalties on a single-tier environment~\cite{TaxonThakur2017, A_Survey_on_Scheduling_2014, AbdelzahirMap2015}. Jia \emph{et al.}~\cite{cacheCloud2013} propose a multi-resource, load balancing greedy algorithm for distributed cloud cache systems. The algorithm seeks a locally optimal schedule of stored data among cache resources so as to minimize the imbalance degree. Resources are allocated priorities/weights according to the system load-distribution, where higher priorities are given to under-utilized resources. Furthermore, a market-based load balancing algorithm is proposed by Yang \emph{et al.}~\cite{marketServer2009} to distribute workloads between resources. The cost of a job is directly related to the resource loads, as well as resources continuously exchange state load-information to decide on the redistribution and allocation of jobs. Heavily utilized resources are assigned higher cost, and thus are not allocated client jobs. The former algorithms minimize the response times of jobs and load imbalance degrees, however they only compute single-tier-driven and penalty-unaware schedules.

The effect of different levels in computational demands and SLA soft deadlines on the system performance of a single-tier environment is investigated by Stavrinides \emph{et al.}~\cite{SaasWork2017}. A tardiness bound relative to the job's service deadline is employed to represent SLA violations. Moon \emph{et al.}~\cite{SoftHardSLA} describe the SLA as a function of response time, where a client's job does not incur an SLA penalty on the cloud service provider if the job completes the execution within pre-defined service bounds. Also, Chen \emph{et al.}~\cite{MinMin_Priority_2013} present a client-priority-aware load balancing algorithm to produce schedules that increase the utilization of resources and reduce the makespan of client jobs. Nayak \emph{et al.}~\cite{NAYAK2018} propose a scheduling mechanism to enhance the acceptance-rate ratio of deadline-sensitive tasks and maximize resource utilization. However, optimization strategies of the former approaches fail to contemplate differentiated QoS penalties for client jobs when the varying levels of SLA violations and tardiness bounds are translated into quantifiable penalties on the cloud service provider. 

Moreover, several scheduling approaches are proposed to improve the latency of client jobs. For instance, the redundancy-based scheduling is a promising reliability approach that makes duplicate copies of a job on multiple resources as presented in Lee \emph{et al.}~\cite{schedRedundant2017}, Birke \emph{et al.}~\cite{partialRedundMulti2017}, and Gardner \emph{et al.}~\cite{Mor2017ReplicRedundant}. Nevertheless, the redundancy approach generally devises scheduling treatment regimes that formulate schedules for client jobs whose QoS penalties are identical, while the various SLA commitments of jobs with their associated differential penalties are not employed when such schedules are produced.

Mailach \emph{et al.}~\cite{schedError2017} schedule jobs based on their estimated service times. Okopa \emph{et al.}~\cite{fixedSchedu2012} present a fixed-priority scheduling to address the execution of client jobs with variant execution demands. The proposed scheduling policy primarily delivers high service performance to high priority jobs by reducing their average response times, nevertheless, it negatively penalizes the performance of low priority jobs. However, such schedules are only single-tier-driven formulated on single-server and multi-server systems, while differential service penalties of jobs are not applied.

Furthermore, pair-based scheduling mechanisms are proposed to minimize the execution time of tasks on resources of multiple clouds~\cite{Panda2018, Pair2018}. Panda \emph{et al.}~\cite{Panda2015} formalize job schedules on multiple clouds, as well as present scheduling algorithms that enhance the makespan of tasks and average utilization of the clouds. One presented scheduling algorithm employs the task minimum completion time as a performance indicator to schedule tasks on the cloud that completes their execution at the earliest time. Another scheduling algorithm computes the median of tasks over all clouds, so as to assign the maximum-median task to the cloud that completes the execution at the earliest opportunity.

Similarly, Moschakis \emph{et al.}~\cite{MOSCHAKIS2015} present a multi-cloud scheduling model, and propose a scheduling strategy to dispatch tasks into the least loaded cloud using an inter-cloud dispatcher. In each cloud, a private cloud dispatcher is employed to distribute incoming tasks with the goal of minimizing the total makespan and maximizing the utilization of resources. However, the former multi-cloud-based scheduling algorithms identically penalize client jobs regardless of the performance effect of their service tardiness and demands on such clouds. Such approaches would in turn fail to formulate optimal schedules when the multiple clouds employ multi-tier environments, primarily when client jobs entail various differentiated SLA penalties to represent their service performance.

Moreover, meta-heuristic approaches are employed to provide near-optimal schedules in a reasonable time \cite{Divya2014heur, evolution2014, Singh2017Areview, MISHRA2018}. Such approaches are typically used because the various characteristics and SLA obligations of clients make tackling the scheduling problem a complex task that often cannot be effectively addressed in a polynomial time. The honey-bee meta-heuristic algorithm has been employed by Babu \emph{et al.}~\cite{Bee_2015, Bee_2016} to distribute workloads between resources and minimize job response times. Jobs removed from overloaded queues are treated as honey bees, while underloaded queues are treated as food sources. Although the honey-bee scheduling approach improves the satisfaction for a specific job, it does not account for the satisfaction status of other client jobs and their effect on the system performance. Thus, other jobs waiting in the queues of the tier would not necessarily be satisfied and benefit from the scheduling decision. 

In addition, Gautam \emph{et al.}~\cite{Guatam2018} use a genetic-based approach to formulate QoS-based optimal schedules that reduce the delay cost of client jobs, however, in a single-tier environment. In a similar environment, a resource scheduling genetic-based approach is proposed by Wang \emph{et al.}~\cite{Wang2013Resource} to allocate independent tasks of known service demands to a set of resources, to minimize the response time and energy consumption cost. Likewise, Boloor \emph{et al.}~\cite{HeuristicSLA2010} present a heuristic-based scheduling approach to tackle the execution of client requests on resources of multiple data centers such that the percentile of requests' response times is less than a pre-defined value. Zhan \emph{et al.}~\cite{Zhan2014Genetic} present a load-balance-aware, genetic-based scheduling method to minimize the makespan of client jobs. Nevertheless, the former meta-heuristic approaches do not address the differentiated penalties of client jobs at the system-metric level of delay cost and response time of client jobs. Also, such schedules produce non-optimal performance when measured at the multi-tier level.

Furthermore, a combination of Ant Colony Optimization (ACO) and Particle Swarm Optimization (PSO) algorithms is presented by Cho \emph{et al.}~\cite{Cho2015} to jointly compose an ACOPS algorithm that balances the load between resources. The PSO operator is used to speedup the convergence procedure of the ACO scheduling. However, the ACOPS employs a pre-reject operator to reject tasks that demand for a memory larger than the remaining memory in the system. Although the pre-reject operator reduces the scheduling solution space and time produced by the ACO algorithm, the various SLA commitments of tasks make such rejection strategies increase QoS penalties and the likelihood of dissatisfied clients. The performance of schedules formulated through such meta-heuristic approaches are not optimized at the tier and multi-tier levels of the cloud computing environment.

Reig \emph{et al.}~\cite{predict2010} rely on prediction models to identify the resource requirements that the client is entitled to consume, so as to avoid inefficient allocation and utilization of resources. However, such models do not distribute the load among resources by means of optimal schedules to guarantee the explicit SLA obligations of clients. Nevertheless, maximizing resource utilization would often incur potential SLA violations, while focusing on satisfying SLA requirements of clients may imply poor resource utilization~\cite{nearOptimal2018}.

Generally speaking, existing approaches adopt identical SLA penalties for jobs that demand for optimal QoS-aware schedules formulated through either single-tier-driven or resource-driven strategies. Because client jobs often tend to have different tolerances and sensitivities to SLA violations, such approaches in their optimization strategies would formulate schedules that do not account for the performance impact of the differential SLA penalties of client jobs at the multi-tier level of the cloud computing environment. An optimal balance should be maintained between meeting QoS obligations specified in the SLA and mitigating commercial penalties associated with potential SLA violations.

As such, a differential penalty is a viable performance metric that should be devised to reflect on the QoS provided to clients. This paper presents an SLA-based management approach that mitigates the effect of a penalty in multi-tier cloud computing environments through differential penalty-driven scheduling. Optimal schedules are formulated with respect to SLA commitments of clients, in the context of various QoS and in compliance with the risk operations of the cloud service provider.

\section{Penalty-Oriented Multi-Tier SLA Centric Scheduling of Cloud Jobs}
\label{sec:probFormal}

A multi-tier cloud computing environment consisting of $N$ sequential tiers is considered:
\begin{equation}
T = \left \{T\!_1,T\!_2,T\!_3,...,T\!_N  \right \}
\end{equation}

Each tier $T\!_j$ employs a set of identical computing resources $R_j$:
\begin{equation}
R_j = \left \{R_{j,1},R_{j,2},R_{j,3},...,R_{j,M}  \right \}
\end{equation}

Each resource $R_{j,k}$ employs a queue $Q_{j,k}$ that holds jobs waiting for execution by the resource. Jobs with different resource computational requirements and QoS obligations are submitted to the environment. It is assumed that these jobs are submitted by different clients and hence are governed by various SLA's. Jobs arrive at the environment in streams. A stream $S$ is a set of jobs:
\begin{equation}
S = \left \{ J_1,J_2,J_3,...,J_l  \right \}
\end{equation}

\begin{figure*}[ht!]
	\centering
    \captionsetup{justification=centering}
		\includegraphics[width=0.95\textwidth]{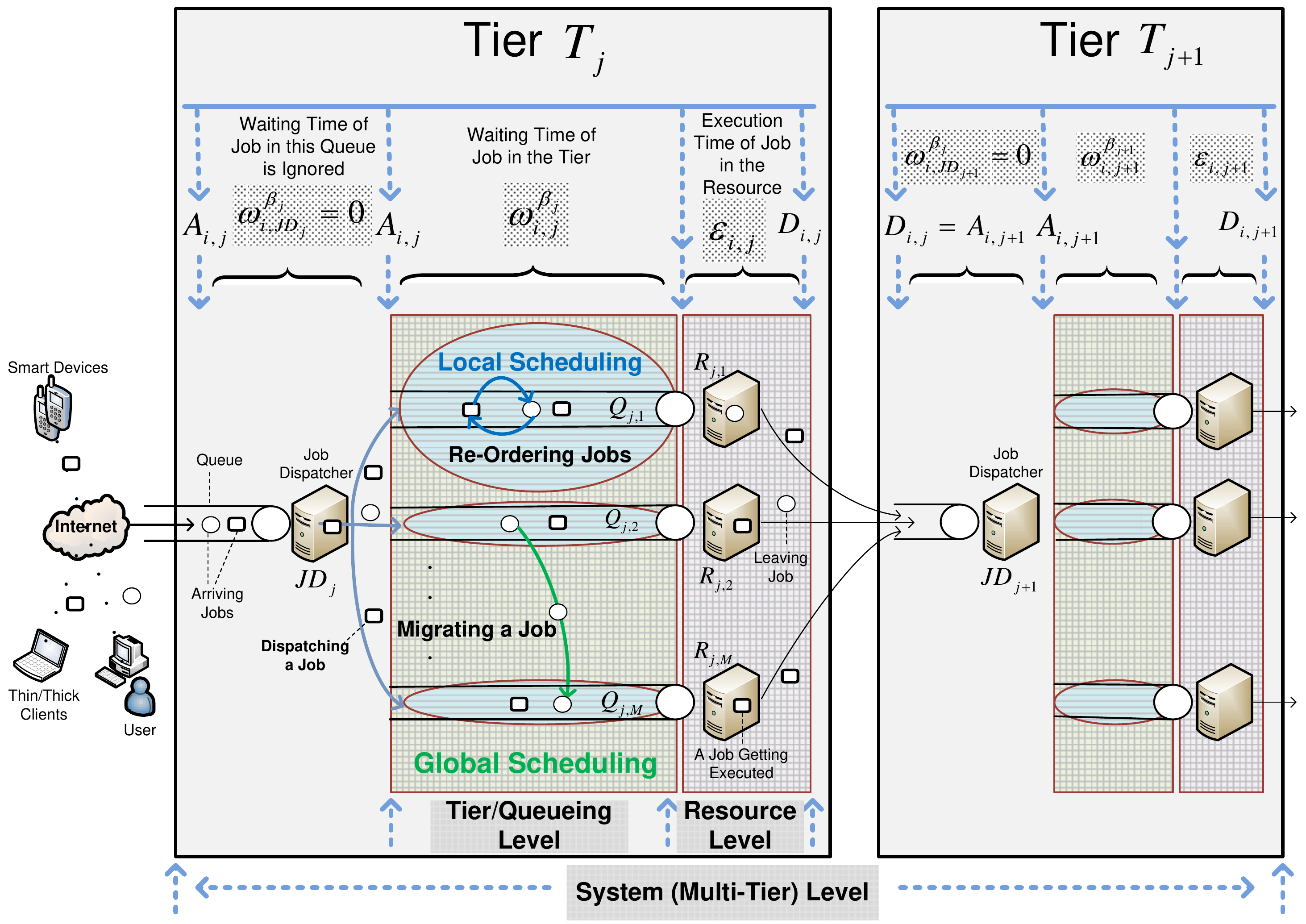}
    \caption{Modeling Parameters and Operators of 2 Consecutive Tiers of the Multi-Tier Cloud Environment}
	\label{fig:systemmodel11}
\end{figure*}

The index of each job $J_i$ signifies its arrival ordering at the environment. For example, job $J_1$ arrives at the environment before job $J_2$. Jobs submitted to tier $T\!_j$ are queued for execution based on an ordering $\beta_j$. As shown in Figure~\ref{fig:systemmodel11}, each tier $T\!_j$ of the environment consists of a set of resources $R_j$. Each resource $R_{j,k}$ has a queue $Q_{j,k}$ to hold jobs assigned to it. For instance, resource $R_{j,1}$ of tier $T\!_j$ is associated with queue $Q_{j,1}$, which consists of $3$ jobs waiting for execution.
\begin{equation}
\label{equ:betaj}
    \beta_j = \bigcup_{k=1}^{M_k} \text{I}(Q_{j,k}),\;\;\;\; \forall j\!\in\![1,N]
\end{equation}
where $\text{I}(Q_{j,k})$ represents indices of jobs in $Q_{j,k}$. For instance, $\nolinebreak{\text{I}(Q_{1,2})=\{3,5,2,7\}}$ signifies that jobs $J_3$, $J_5$, $J_2$, and $J_7$ are queued in $Q_{1,2}$ such that job $J_3$ precedes job $J_5$, which in turn precedes job $J_2$, and so on.

Jobs arrive in random manner. A job dispatcher $J\!D_j$ is employed to buffer incoming client jobs to tier $T\!_j$. Job $J_i$ arrives at tier $T\!_j$ at time $A_{i,j}$ via the queue of the job dispatcher $J\!D_j$ of the tier. It has a prescribed execution time $\mathcal{E}_{i,j}$ at each tier. Each job has a service deadline $\mathcal{D\!L}_i$, which in turn stipulates a target completion time $\mathcal{C}_i^{(t)}$ for the job $J_i$ in the multi-tier environment.
\begin{equation}
\label{equ:Ji}
    J_i = \left \{A_{i,j}, \mathcal{E}_{i,j}, \mathcal{C}_i^{(t)} \right \},\;\;\; \forall\; T\!\!_j\!\in\!T
\end{equation}

The total execution time $\mathcal{E\!T}\!\!_{i}$ of each job $J_i$ is as follows:
\begin{equation}
\label{equ:Ei}
    \mathcal{E\!T}\!\!_i = \sum_{j=1}^{N} \mathcal{E}_{i,j}
\end{equation}

The job dispatcher $J\!D_j$ queues these jobs to the resource queues $R_j$ of the tier. Job $J_{i}$ waits $\omega_{i,j}^{\beta_j}$ time units in tier $T\!_j$ according to an ordering $\beta_j$ of the jobs waiting for execution at resources $R_j$. Job $J_{i}$ gets its turn of execution by resource $R_{j,k}$, and afterward, leaves tier $T\!_j$ at time $D_{i,j}$ to be queued by the dispatcher $J\!D_{\nolinebreak{j+1}}$ of tier $T_{j+1}$. When leaving the cloud environment from tier $N$, job $J_i$ has a response time $\mathcal{R\!T}_{\!\!\!i}^\beta$ and end-to-end waiting time $\omega\!\mathcal{T}_i^\beta$ computed according to the overall ordering $\beta$ of jobs at the $N$ tiers.
\begin{equation}
\label{equ:beta}
    \beta = \bigcup_{j=1}^{N} \beta_j
\end{equation}

The waiting time $\omega_{i,j}^{\beta_j}$ of each job $J_i$ at tier $T\!_j$ is defined as the difference between the time it starts execution by one of the resources and its arrival time $A_{i,j}$. The end-to-end waiting time $\omega\!\mathcal{T}_i^\beta$ of job $J_i$ according to the overall ordering $\beta$ across all tiers in the multi-tier cloud environment is defined as the summation of the job's waiting time $\omega_{i,j}^{\beta_j}$ in all tiers. The response time $\mathcal{R\!T}_{\!\!\!i}^\beta$ of job $J_i$ in the multi-tier cloud environment is defined as the difference between the departure time $D_{i,N}$ of job $J_i$ from the last tier $T\!_N$ and the arrival time $A_{i,1}$ of job $J_i$ to the first tier $T\!_1$. The response time $\mathcal{R\!T}_{\!\!\!i}^\beta$ of job $J_i$ can also be viewed as the summation of waiting times $\omega_{i,j}^{\beta_j}$ and execution times $\mathcal{E}_{i,j}$. The performance parameters $\omega_{i,j}^{\beta_j}$, $\omega\!\mathcal{T}_i^\beta$, and $\mathcal{R\!T}_{\!\!\!i}^\beta$ for each job $J_i$ are computed as follows:
\begin{equation}
\label{equ:W}
    \omega_{i,j}^{\beta_j} = D_{i,j} - \mathcal{E}_{i,j} - A_{i,j}
\end{equation}
\begin{equation}
\label{equ:TW}
    \omega\!\mathcal{T}_i^\beta = \sum_{j=1}^{N} \omega_{i,j}^{\beta_j}
\end{equation}
\begin{equation}
\label{equ:TR}
    \mathcal{R\!T}_{\!\!\!i}^\beta = D_{i,N} - A_{i,1} = \sum_{j=1}^{N} (\omega_{i,j}^{\beta_j} + \mathcal{E}_{i,j}) = \omega\!\mathcal{T}_i^\beta + \mathcal{E\!T}\!\!_i
\end{equation}

The excessive volume of client demands and the potential lack of adequate resource availability are critical situations for the cloud service providers. Priorities are, therefore, given to jobs according to the impact of potential delays in their execution. Such priorities must be reflected in the scheduling strategy in a way that ensures the financial viability of the cloud service provider and, at the same time, high client satisfaction. The scheduling strategy should leverage the available delay tolerance of client jobs so as to satisfy the critical demands of delay intolerant jobs.

\subsection{Differentiated Cost of Time-Based Scheduling}
\label{sec:waitingFinancialPenalty}

The execution time $\mathcal{E}_{i,j}$ of job $J_i$ at tier $T\!_j$ is pre-defined in advance. Therefore, the resource capabilities of each tier $T_j$ are not considered and, thus, the total execution time $\mathcal{E\!T}\!\!_i$ of job $J_i$ is constant. Instead, the primary concern is on the queueing-level of the environment represented by the total waiting time $\omega\!\mathcal{T}_i^{\beta}$ of job $J_i$ at all tiers $T$ according to the ordering $\beta$.

A unit of waiting time $\omega\!\mathcal{T}\!\!{_i}$ of job $J_i$ would incur a differentiated financial service cost $\psi_i$. Such situations demand the cloud service provider emphasize the notion of financial penalty in the scheduling of client jobs so that schedules are computed based on economic considerations. The service penalty cost $\psi_i$ is assumed to follow a normal distribution with a mean $\mu$ and variance $\sigma$.
\begin{equation}
\label{equ:psi}
\begin{split}
    \psi_i & = N(\mu,\sigma)
\end{split}
\end{equation}

The service time of job $J_i$ is subject to an SLA that stipulates an exponential differentiated financial penalty curve $\eta_i$ as follows:
\begin{equation}
\label{equ:etai1}
\begin{split}
    \eta_i & = \chi * ( 1 - \text{e}^{- \nu \; \psi_i \; \sum_{j=1}^{N} \omega_{i,j}^{\beta_j} } )
\end{split}
\end{equation}

As such, the total differentiated financial performance penalty cost of the job stream $l$ across all tiers is given by $\vartheta$ as follows:
\begin{equation}
\label{equ:vartheta1}
    \vartheta = \sum_{i=1}^{l} \eta_i
\end{equation}

The objective is to find job orderings $\nolinebreak{\beta=(\beta_1, \beta_2, \beta_3, \ldots, \beta_N)}$ such that the stream's total differentiated financial penalty cost $\vartheta$ is minimal:
\begin{equation}
\label{equ:minVARTHETA1}
\begin{split}
     \underset{\beta}{\text{minimize}} \; (\vartheta) & \equiv \underset{\beta}{\text{minimize}} \; \sum_{i=1}^{l} \sum_{j=1}^{N} \; (\; \psi_i \; \omega_{i,j}^{\beta_j} \;)
\end{split}
\end{equation}

\subsection{Differentiated Cost of Time-Based Scheduling: Multi-Tier Considerations}
\label{sec:SLAFinancialPenalty}

The target completion time $\mathcal{C}_i^{(t)}$ of job $J_i$ represents an explicit QoS obligation on the service provider to complete the execution of the job. Thus, the $\mathcal{C}_i^{(t)}$ incurs a service deadline $\mathcal{D\!L}_i$ for the job in the environment. The service deadline $\mathcal{D\!L}_i$ is higher than the total prescribed execution time $\mathcal{E\!T}\!\!_i$ and incurs a total waiting time allowance $\omega\!\mathcal{A\!L}_i$ for job $J_i$ in the environment.
\begin{equation}
\label{equ:DLi}
\begin{split}
   \mathcal{D\!L}_i & = \mathcal{C}_i^{(t)} - A_{i,j} \\
                    & = \mathcal{E\!T}\!\!_i + \omega\!\mathcal{A\!L}_i
\end{split}
\end{equation}

As such, the time difference between the response time $\mathcal{R\!T}_{\!\!\!i}^\beta$ and the service deadline $\mathcal{D\!L}_i$ represents the service-level violation time $\alpha_i^\beta$ of job $J_i$, according to the ordering $\beta$ of jobs in tiers $T$ of the environment.
\begin{equation}
\label{equ:alpha_i}
(\mathcal{R\!T}_{\!\!\!i}^\beta - \mathcal{D\!L}_i) =
\begin{cases}
     \alpha_i^\beta > 0, \;\; \text{The client is not satisfied}  \\
     \alpha_i^\beta \leq 0, \;\; \text{The client is satisfied}
\end{cases}
\end{equation}

A unit of SLA violation time $\alpha_i^\beta$ of the job $J_i$ at the multi-tier level of the environment incurs a differentiated financial SLA violation cost $\zeta_i$. The cost $\zeta_i$ of SLA violation at the multi-tier level is assumed to follow a normal distribution with a mean $\mu$ and variance $\sigma$.
\begin{equation}
\label{equ:zetai}
\begin{split}
    \zeta_i & = N(\mu,\sigma)
\end{split}
\end{equation}

The service-level violation time $\alpha_i^\beta$ is subject to an SLA that stipulates an exponential differentiated financial penalty curve $\eta_i$ as follows:
\begin{equation}
\label{equ:etai2}
\begin{split}
    \eta_i & = \chi * (   1 - \text{e}^{- \nu \; \zeta_i \; (\mathcal{R\!T}_{\!\!i}^\beta - \mathcal{D\!L}_i) }   )\\
           & = \chi * (   1 - \text{e}^{- \nu \; \zeta_i \; (\omega\!\mathcal{T}_i^\beta - \omega\!\mathcal{A\!L}_i) }) \\
           & = \chi * (   1 - \text{e}^{- \nu \; \zeta_i \; \alpha_i^\beta }   )
\end{split}
\end{equation}
where $\chi$ is a monetary cost factor and $\nu$ is an arbitrary scaling factor. The total performance penalty cost $\vartheta$ of the stream $l$ across all tiers is given by Equation~\ref{equ:vartheta1} and, accordingly, the financial performance of job schedules is optimized such that the differentiated SLA violation penalty is minimized at the multi-tier level.

\begin{itemize}
  \item \emph{Differentiated $\omega\!\mathcal{A\!L}_i$ Based Minimum Penalty Formulation}
\end{itemize}

The performance of job schedules is formulated with respect to the multi-tier waiting time allowance $\omega\!\mathcal{A\!L}_i$ of each job $J_i$. Accordingly, the SLA violation penalty is evaluated at the multi-tier level of the environment. The objective is to seek job schedules in tiers of the environment such that the total SLA violation penalty of jobs would be minimized \emph{globally} at the multi-tier level of the environment.

The total waiting time $\omega\!\mathcal{T}^\beta_i$ of job $J_i$ currently waiting in tier $T\!_p$, where $\nolinebreak{p\!<\!N}$, is not totally known because the job has not yet completely finished execution from the multi-tier environment. Therefore, the job's $\omega\!\mathcal{T}^\beta_i$ at tier $T\!_p$ is estimated and, thus, represented by $\omega\mathcal{C\!X}_{i,p}^\beta$ according to the scheduling order $\beta$ of jobs. As such, the job's service-level violation time $\alpha_i^{\beta}$ at tier $T\!_p$ would be represented by the expected waiting time $\omega\mathcal{C\!X}_{i,p}^\beta$ of job $J_i$ in the current tier $T\!_p$ and the waiting time allowance $\omega\!\mathcal{A\!L}_i$ incurred from the job's service deadline $\mathcal{D\!L}_i$ at the multi-tier level of the environment.
\begin{equation}
\label{equ:alpha_i}
\begin{split}
   \alpha_i^\beta & = \omega\mathcal{C\!X}_{i,p}^\beta - \omega\!\mathcal{A\!L}_i
\end{split}
\end{equation}
where the expected waiting time $\omega\mathcal{C\!X}_{i,p}^\beta$ of job $J_i$ at tier $T\!_p$ incurs the total waiting time $\omega\!\mathcal{T}_i^\beta$ of job $J_i$ at the multi-tier level.
\begin{equation}
\label{equ:wEip}
\begin{split}
   \omega\mathcal{C\!X}_{i,p}^\beta & = \sum_{j=1}^{(p-1)}(\omega_{i,j}^{\beta_j}) + \omega\!E\!\mathcal{L}_{i,p} + \omega\!\mathcal{R\!M}_{i,p}^{\beta_p}
\end{split}
\end{equation}
where $\nolinebreak{\omega_{i,j}^{\beta_j}(\forall j\leq(p-1))}$ represents the waiting time of job $J_i$ in each tier $T\!_j$ in which the job has completed execution, $\nolinebreak{\omega\!E\!\mathcal{L}_{i,p}}$ represents the elapsed waiting time of job $J_i$ in the tier $T\!_p$ where the job currently resides, and $\nolinebreak{\omega\!\mathcal{R\!M}_{i,p}^{\beta_p}}$ represents the remaining waiting time of job $J_i$ according to the scheduling order $\beta_p$ of jobs in the current holding tier $T\!_p$.
\begin{equation}
\label{equ:wijbetaj}
    \omega\!\mathcal{R\!M}_{i,j}^{\beta_j} = \sum_{h\in \text{I}(Q_{j,k}),\;h\;\text{precedes job}\;J_i}^{\forall} \mathcal{E}_{h,j},\;\;\;\; \forall j\!\in\![1,N]
\end{equation}
where $\text{I}(Q_{j,k})$ represents indices of jobs in $Q_{j,k}$. For instance, $\nolinebreak{\text{I}(Q_{1,2})=\{3,5,2,7\}}$ signifies that jobs $J_3$, $J_5$, $J_2$, and $J_7$ are queued in $Q_{1,2}$ such that job $J_3$ precedes job $J_5$, which in turn precedes job $J_2$, and so on. However, the elapsed waiting time $\omega\!E\!\mathcal{L}_{i,j}$ affects the execution priority of the job. The higher the time of $\omega\!E\!\mathcal{L}_{i,j}$ of job $J_i$ in the tier $T\!_j$, the lower the remaining allowed time of $\omega\!\mathcal{A\!L}_i$ of job $J_i$ at the multi-tier level, thus, the higher the execution priority of job $J_i$ in the resource.

The objective is to find scheduling orders $\nolinebreak{\beta=(\beta_1, \beta_2, \beta_3, \ldots, \beta_N)}$ for jobs of each tier $T\!_j$ such that the stream's total differentiated penalty $\vartheta$ is minimal, and thus the SLA violation penalty is minimal. The financially optimal performance scheduling with respect to $\omega\!\mathcal{A\!L}_i$ is formulated as:
\begin{equation}
  \label{equ:min_3}
    \begin{split}
     \underset{\beta}{\text{minimize}} \; (\vartheta) & \equiv \underset{\beta}{\text{minimize}} \sum_{i=1}^{l}\sum_{p=1}^{N} \zeta_i \; (\omega\mathcal{C\!X}_{i,p}^\beta - \omega\!\mathcal{A\!L}_i)
    \end{split}
\end{equation}

\begin{itemize}
  \item \emph{Differentiated $\omega\!\mathcal{P\!T}\!\!_{i,j}$ Based Minimum Penalty Formulation}
\end{itemize}

The performance of job schedules is formulated with respect to a differentiated waiting time $\omega\!\mathcal{P\!T}\!\!_{i,j}$ of the job $J_i$ at each tier $T\!_j$. The $\omega\!\mathcal{P\!T}\!\!_{i,j}$ is derived from the multi-tier waiting time allowance $\omega\!\mathcal{A\!L}_i$ of job $J_i$, with respect to the execution time $\mathcal{E}_{i,j}$ of the job $J_i$ at the tier level relative to the job's total execution time $\mathcal{E\!T}\!\!_i$ at the multi-tier level of the environment.
\begin{equation}
\label{equ:wPT}
\begin{split}
     \omega\!\mathcal{P\!T}\!_{i,j} = \omega\!\mathcal{A\!L}_i * \frac{\mathcal{E}_{i,j}}{\mathcal{E\!T}\!\!_i}
\end{split}
\end{equation}

In this case, the higher the execution time $\mathcal{E}_{i,j}$ of job $J_i$ in tier $T\!_j$, the higher the job's differentiated waiting time allowance $\omega\!\mathcal{P\!T}\!_{i,j}$ in the tier $T\!_j$. Accordingly, the SLA violation penalty is evaluated at the multi-tier level with respect to the $\omega\!\mathcal{P\!T}\!\!_{i,j}$ of each job $J_i$.

The waiting time $\omega_{i,j}^{\beta_j}$ of job $J_i$ at tier $T\!_j$ would not be totally known until the job completely finishes execution from the tier, however, it can be estimated by $\omega\!\mathcal{P\!X}_{i,j}^{\beta_j}$ according to the current scheduling order $\beta_j$ of jobs in the tier $T\!_j$. As such, the service-level violation time $\alpha\!\mathcal{T}_{i,j}^{\beta_j}$ of job $J_i$ in the tier $T\!_j$ according to the scheduling order $\beta_j$ of jobs would be represented by the expected waiting time $\omega\!\mathcal{P\!X}_{i,j}^{\beta_j}$ and the differentiated waiting time allowance $\omega\!\mathcal{P\!T}\!\!_{i,j}$, of the job in the tier $T\!_j$.
\begin{equation}
\label{equ:alphaT_ij}
\begin{split}
   \alpha\!\mathcal{T}_{i,j}^{\beta_j} & = \omega\!\mathcal{P\!X}_{i,j}^{\beta_j} - \omega\!\mathcal{P\!T}\!\!_{i,j}
\end{split}
\end{equation}
\begin{equation}
\label{equ:alphaT_ij}
\begin{split}
   \alpha_{i}^{\beta} & = \sum_{j=1}^{N} \alpha\!\mathcal{T}_{i,j}^{\beta_j}
\end{split}
\end{equation}
where $\alpha_{i}^{\beta}$ is the total service-level violation time of the job $J_i$ at all tiers of the environment according to the scheduling order $\beta$. The expected waiting time $\omega\!\mathcal{P\!X}_{i,j}^{\beta_j}$ incurs the actual waiting time $\omega_{i,j}^{\beta_j}$ of job $J_i$ in tier $T\!_j$, and thus depends on the elapsed waiting time $\nolinebreak{\omega\!E\!\mathcal{L}_{i,j}}$ and the remaining waiting time $\nolinebreak{\omega\!\mathcal{R\!M}_{i,j}^{\beta_j}}$ of the job $J_i$ according to the scheduling order $\beta_j$ of jobs in the current holding tier $T\!_j$.
\begin{equation}
\label{equ:wCXij}
\begin{split}
   \omega\!\mathcal{P\!X}_{i,j}^{\beta_j} & = \omega\!E\!\mathcal{L}_{i,j} + \omega\!\mathcal{R\!M}_{i,j}^{\beta_j}
\end{split}
\end{equation}

The elapsed waiting time parameter $\omega\!E\!\mathcal{L}_{i,j}$ of job $J_i$ in tier $T\!_j$ affects the job's execution priority in the resource. The higher the time of $\omega\!E\!\mathcal{L}_{i,j}$, the lower the remaining time of the differentiated waiting allowance $\omega\!\mathcal{P\!T}\!\!_{i,j}$ of job $J_i$ in the tier $T\!_j$, therefore, the higher the execution priority of the job $J_i$ in the resource, so as to reduce the service-level violation time $\alpha\!\mathcal{T}_{i,j}^{\beta_j}$ of the job in the tier $T\!_j$.

The objective is to find scheduling orders $\nolinebreak{\beta=(\beta_1, \beta_2, \beta_3, \ldots, \beta_N)}$ for jobs of each tier $T\!_j$ such that the stream's total differentiated penalty $\vartheta$ is minimal, and thus the SLA violation penalty is minimal. The financially optimal performance scheduling with respect to $\omega\!\mathcal{P\!T}\!\!_{i,j}$ is formulated as:
\begin{equation}
    \label{equ:min_4}
    \begin{split}
     \underset{\beta}{\text{minimize}} \; (\vartheta) & \equiv \underset{\beta}{\text{minimize}} \sum_{i=1}^{l} \sum_{j=1}^{N} \zeta_i \; (\omega\!\mathcal{P\!X}_{i,j}^{\beta_j} - \omega\!\mathcal{P\!T}\!_{i,j})
    \end{split}
\end{equation}

\section{Minimum Penalty Job Scheduling: A Genetic Algorithm Formulation}
\label{sec:penaltyAppr}

During scheduling of client jobs for execution, a job is first submitted to tier-1 by one of the resources of the tier. Jobs should be scheduled in such a way that minimizes total waiting time and SLA-violation penalties. Finding a job scheduling that yields minimum penalty is an NP problem. Given the expected volume of jobs to be scheduled and the computational complexity of the job scheduling problem, it is prohibitive to seek optimal solution for the job scheduling problem using exhaustive search techniques. Thus, a meta-heuristic search strategy, such as Permutation Genetic Algorithms (PGA), is a viable option for exploring and exploiting the large space of scheduling permutations~\cite{GA3}. Genetic algorithms have been successfully adopted in various problem domains~\cite{GaTabu1}, and have undisputed success in yielding near optimal solutions for large scale problems, in reasonable time~\cite{heuristicPSO}.

Scheduling client jobs entails two steps: (1) allocating/distributing the jobs among the different tier resources. Jobs that are allocated to a given resource are queued in the queue of that resource; (2) ordering the jobs in the queue of the resource such that their penalty is minimal. What makes the problem increasingly hard is the fact that jobs continue to arrive, while the prior jobs are waiting in their respective queues for execution. Thus, the scheduling process needs to respond to the job arrival dynamics to ensure that job execution at all tiers is penalty optimal. To achieve this, job ordering in each queue should be treated as a continuous process. Furthermore, jobs should be migrated from one resource to another so as to ensure balanced job allocation and maximum resource utilization. Thus, two operators are employed for constructing optimal job schedules:
\begin{itemize}
  \item The \emph{reorder} operator is used to change the ordering of jobs in a given queue so as to find an order that minimizes the total penalty of all jobs in the queue.
  \item The \emph{migrate} operator, in contrast, is used to exploit the benefits of moving jobs between the different resources of the tier so as to reduce the total penalty. This process is adopted at each tier of the environment.
\end{itemize}

However, implementing the \emph{reorder/migrate} operators in a PGA search strategy is not a trivial task. This implementation complexity can be relaxed by virtualizing the queues of each tier into one virtual queue. The virtual queue is simply a cascade of the queues of the resources. In this way, the two operators are converged into simply a reorder operator. Furthermore, this simplifies the PGA solution formulation. A consequence of this abstraction is the length of the permutation chromosome and the associated computational cost. This virtual queue will serve as the chromosome of the solution. An index of a job in this queue represents a gene. The ordering of jobs in a virtual queue signifies the order at which the jobs in this queue are to be executed by the resource associated with that queue. Solution populations are created by permuting the entries of the virtual queue, using the \emph{order} and \emph{migrate} operators.
\begin{figure*}[!ht]
\centering
\captionsetup{justification=centering}
	  \includegraphics[width=0.45\textwidth,height=0.342\textheight]{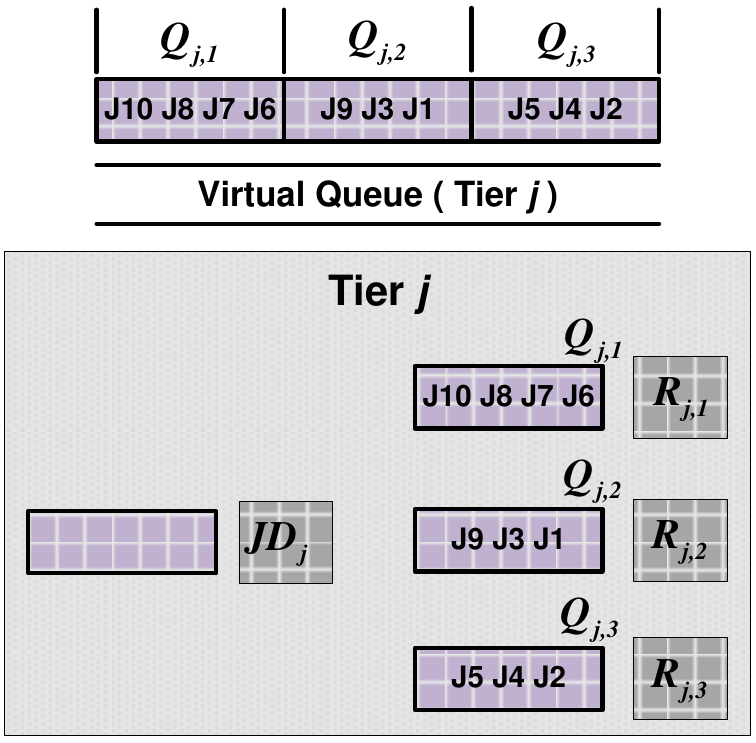}
	  \caption{The Virtual Queue of a Tier $j$}
      \label{fig:tierVirtualQueue}
\end{figure*}
\begin{figure*}[!ht]
\centering
\captionsetup{justification=centering}
	  \includegraphics[width=0.8\textwidth,height=0.656\textheight]{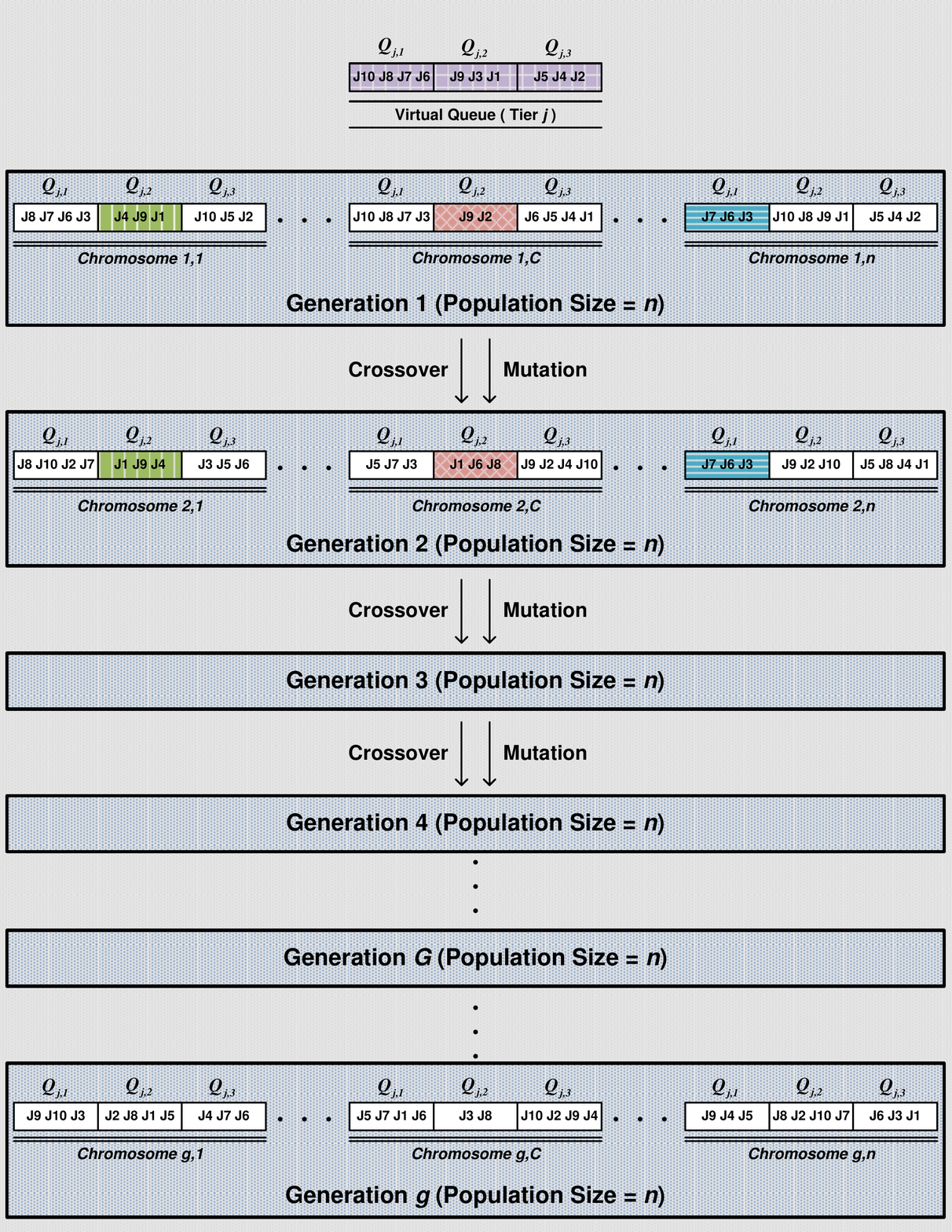}
	  \caption{A Tier-based Genetic Approach on the Virtual Queue}
      \label{fig:GA}
\end{figure*}

\subsection{Tier-Based Virtual Queue}
\label{sec:tierBasedVQ}

To produce tier-driven optimal performance, a tier-based virtual queue is proposed. In this case, a virtual queue is a cascade of resource queues of the tier. Figures~\ref{fig:tierVirtualQueue} and \ref{fig:GA} of the $j^{\text{th}}$ tier show the construct of one virtual queue represented as a cascade of the three queues ($Q_{j,1}$, $Q_{j,2}$, and $Q_{j,3}$) of the tier. The schedule's performance is optimized at this virtual-queue level. 

\subsubsection{Evaluation of Schedules}
\label{sec:step3}

A fitness evaluation function is used to assess the quality of each virtual-queue realization (chromosome). The fitness value of the chromosome captures the cost of a potential schedule. The fitness value $f_{r,G}$ of a chromosome $r$ in generation $G$ is represented by the differentiated financial waiting penalty of the job schedule in the virtual queue, according to the scheduling order $\beta_j$ of jobs in each tier $T_j$.
\begin{equation}
  \label{equ:fitness3}
  f_{r,G} = \sum_{i=1}^{l} (\psi_i \; \omega_{i,j}^{\beta_j})
\end{equation}

The waiting time $\omega_{i,j}^{\beta_j}$ of the $i^{\text{th}}$ job in the virtual queue of the $j^{\text{th}}$ tier should be calculated based on its order in the queue, as per the ordering $\beta_j$. The normalized fitness value $F_r$ of each schedule candidate is computed as follows:
\begin{equation}
  \label{equ:probabilityRouletteWheel}
  F_r  = \frac{f_{r,G}}{\sum_{C=1}^{n} (f_{C,G})}\;,\;\;\;r\!\in\!C
\end{equation}
\begin{figure*}[!ht]
\centering
\captionsetup{justification=centering}
	  \includegraphics[width=\textwidth]{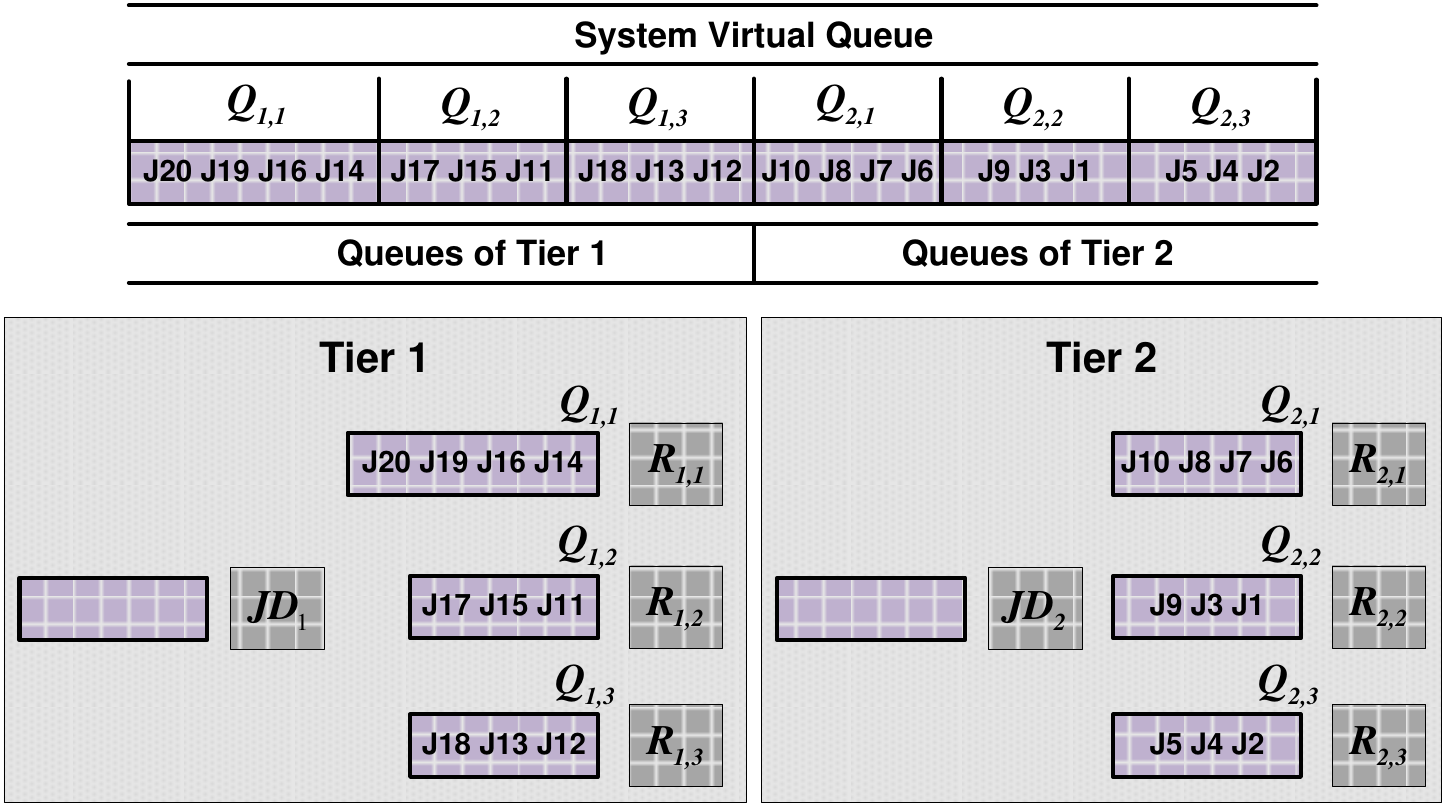}
	  \caption{The System Virtual Queue}
      \label{fig:systemVirtualQueue}
\end{figure*}

Based on the normalized fitness values of the candidates, Russian Roulette is used to select a set of schedule candidates to produce the next generation population, using the combination and mutation operators.

\subsubsection{Evolving the Scheduling Process}
\label{sec:step4}

To evolve a new population that holds new scheduling options for jobs in resource queues of the tier, the crossover and mutation genetic operators are both applied on randomly selected schedules (virtual queues) of the current generation. The crossover operator produces a new generation of virtual queues from the current generation. The mutation operator applies random changes on a selected set of virtual queues of the new generation to produce altered virtual queues. These operators diversify the search direction into new search spaces to avoid getting stuck in a locally optimum solution. Overall, the \emph{Single-Point} crossover and \emph{Insert} mutation genetic operators are used. Rates of crossover and mutation operators are both set to $0.1$ of the population size in each generation.

Figure~\ref{fig:GA} explains how each virtual queue in a given generation is evolved to create a new virtual queue of the next generation, using the crossover and mutation operators. Each chromosome (virtual queue) represents a new scheduling of jobs. The jobs and their order of execution on the resource will be reflected by the segment of the virtual queue corresponding to the actual queue associated with the resource. As a result of the evolution process, each segment of the virtual queue corresponding to an actual queue will be in one of the following states:
\begin{itemize}
  \item Maintain the same set and order of jobs held in the previous generation;
  \item Get a new ordering for the same set of jobs held in the previous generation;
  \item Get a different set of jobs and a new ordering.
\end{itemize}

For instance, queue $Q_{j,1}$ of \nolinebreak{\emph{Chromosome} ($1,\!n$)} in the first generation maintains exactly the same set and order of jobs in the second generation shown in queue $Q_{j,1}$ of \nolinebreak{\emph{Chromosome} ($2,\!n$)}. In contrast, queue $Q_{j,2}$ of \nolinebreak{\emph{Chromosome} ($1,\!1$)} in the first generation maintains the same set of jobs in the second generation, yet has got a new order of jobs as shown in queue $Q_{j,2}$ of \nolinebreak{\emph{Chromosome} ($2,\!1$)}. Finally, queue $Q_{j,2}$ of a random \nolinebreak{\emph{Chromosome} ($1,\!C$)} in the first generation has neither maintained the same set nor the same order of jobs in the second generation shown in queue $Q_{j,2}$ of \nolinebreak{\emph{Chromosome} ($2,\!C$)}, which in turn would yield a new scheduling of jobs in the queue of resource $R_{j,2}$ if \nolinebreak{\emph{Chromosome} ($2,\!C$)} is later selected as the best chromosome of the tier-based genetic solution.

\subsection{Multi-Tier-Based Virtual Queue}
\label{sec:MultitierBasedVQ}

The goal is to formulate optimal schedules such that SLA violation penalties of jobs are reduced at the multi-tier level. However, it is complicated to apply the allocation and ordering operators at the multi-tier level. As such, the operator complexities are mitigated by virtualizing resource queues of the multi-tier environment into a single system virtual queue that represents the chromosome of the scheduling solution, as shown in Figure~\ref{fig:systemVirtualQueue}. This system-level abstraction converges the operators into simply a reorder operator running at the multi-tier level.
\begin{figure*}[!t]
\centering
\captionsetup{justification=centering}
	  \includegraphics[width=\textwidth]{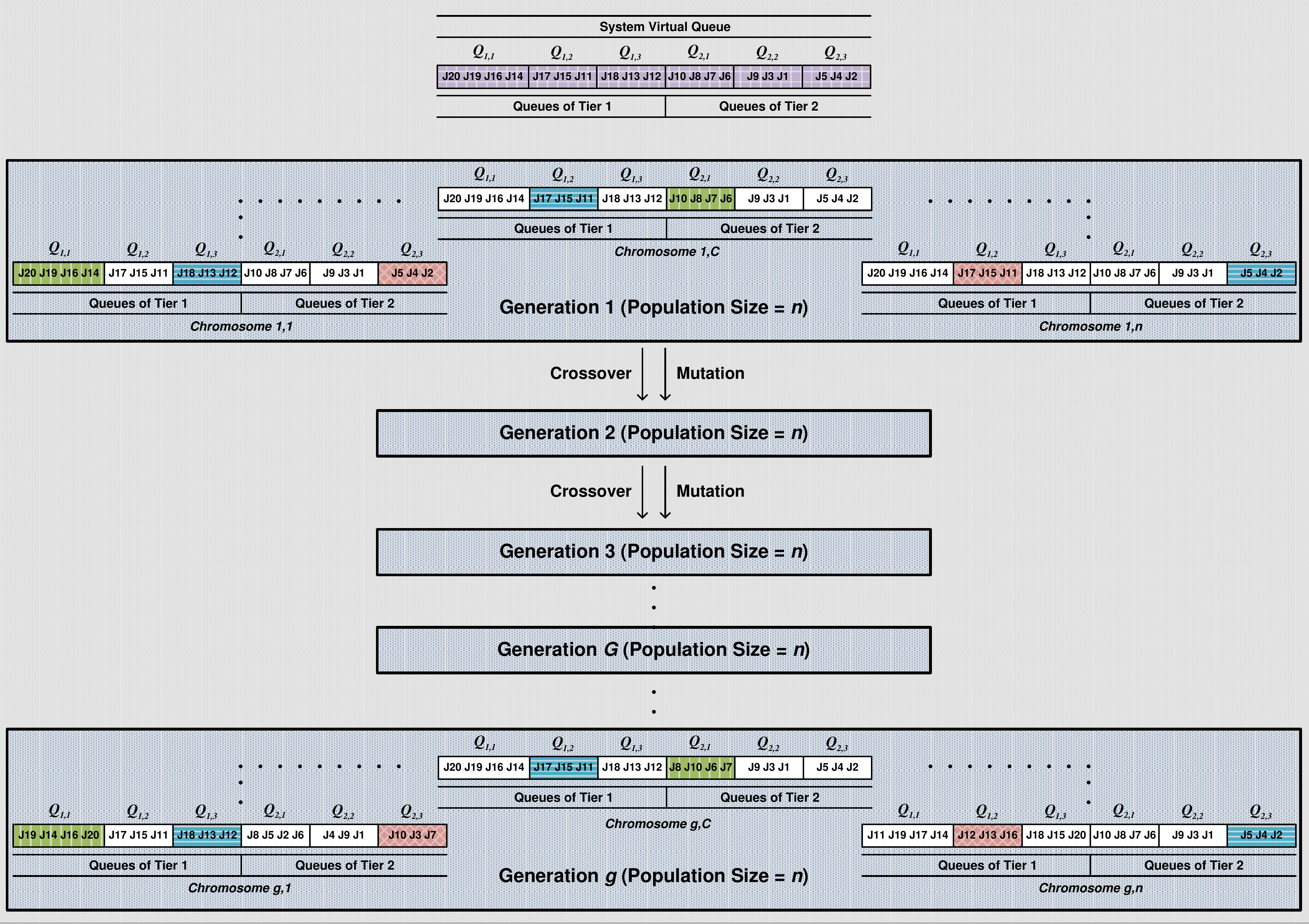}
	  \caption{A System Virtualized Queue Genetic Approach}
      \label{fig:GA2}
\end{figure*}

\subsubsection{Evaluation of Schedules}
\label{sec:step33}

The quality of a job schedule in a system virtual queue realization (chromosome) is assessed by a fitness evaluation function. For a chromosome $r$ in generation $G$, the fitness value $f_{r,G}$ is represented by the SLA violation cost of the schedule in the system virtual queue computed at the multi-tier level. Two different fitness evaluation functions are adopted in two different solutions:
\begin{equation}
\label{equ:fitness4}
f_{r,G} = \\
\begin{cases}
     \sum_{i=1}^{l} \zeta_i \; (\omega\mathcal{C\!X}_{i,p}^\beta - \omega\!\mathcal{A\!L}_i), \;\\ \;\;\;\;\; ~\text{Differentiated Penalty}~\omega\!\mathcal{A\!L}_i~\text{based Scheduling}  \\
     \sum_{i=1}^{l} \zeta_i \; (\omega\!\mathcal{P\!X}_{i,j}^{\beta_j} - \omega\!\mathcal{P\!T}\!_{i,j}), \;\\ \;\;\;\;\;\; \text{Differentiated Penalty}~\omega\!\mathcal{P\!T}\!_{i,j}~\text{based Scheduling}
\end{cases}
\end{equation}

In both scenarios, the SLA violation cost of job $J_i$ is represented by the job's waiting time (either $\omega\mathcal{C\!X}_{i,p}^\beta$ or $\omega\!\mathcal{P\!X}_{i,j}^{\beta_j}$) according to its scheduling order $\beta$ in the system virtual queue and the job's waiting allowance (either $\omega\!\mathcal{A\!L}_i$ or $\omega\!\mathcal{P\!T}\!_{i,j}$) incurred from its service deadline $\mathcal{D\!L}_i$ at the multi-tier level.
The normalized fitness value $F_r$ of each schedule candidate is computed as in Equation~\ref{equ:probabilityRouletteWheel}. Based on the normalized fitness values of the candidates, Russian Roulette is used to select a set of schedule candidates that produce the next generation population, using the combination and mutation operators.

\subsubsection{Evolving the Scheduling Process}
\label{sec:step44}

The schedule of the system virtual queue is evolved to produce a population of multiple system virtual queues, each of which represents a chromosome that holds a new scheduling order of jobs at the multi-tier level. To produce a new population, the \emph{Single-Point} crossover and \emph{Insert} mutation genetic operators are applied on randomly selected system virtual queues from the current population. Rates of these operators in each generation are set to be $0.1$ of the population size. The evolution process of schedules of the system virtual queues along with the genetic operators are explained in Figure~\ref{fig:GA2}. Each segment in the system virtual queue corresponds to an actual queue associated with a resource in the tier. In each generation, each segment is subject to the states examined in Section~\ref{sec:step4}.

\section{Experimental Work and Discussions on Results}
\label{sec:expWork3}

The tier-based and multi-tier-based differentiated SLA-driven penalty scheduling are applied on the multi-tier environment. The differentiated service penalty cost $\psi_i$ and SLA violation cost $\zeta_i$ for each job are generated using a mean $\mu$ of $1,\!000$ cost units and a variance $\sigma$ of $25$. The penalty parameter $\nu$ is set to $\nolinebreak{\nu\!=\!\frac{0.01}{1000}}$.

\begin{figure*}[!ht]
        \centering
        \captionsetup{justification=centering}
        \begin{subfigure}{0.48875\textwidth}
		        \includegraphics[width=\textwidth]
                {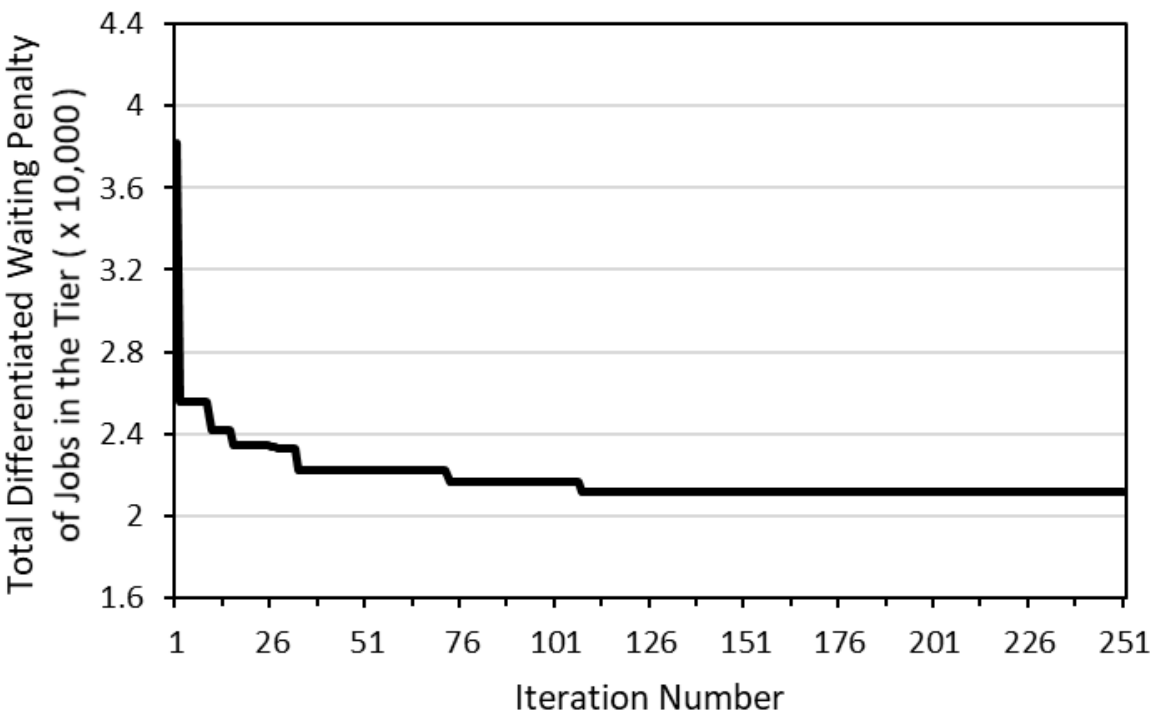}
	            \caption{Virtual Queue of 15 Jobs}
	            \label{fig:TierBased_38203}
        \end{subfigure}
        ~
        \begin{subfigure}{0.48875\textwidth}
                \includegraphics[width=\textwidth]
                {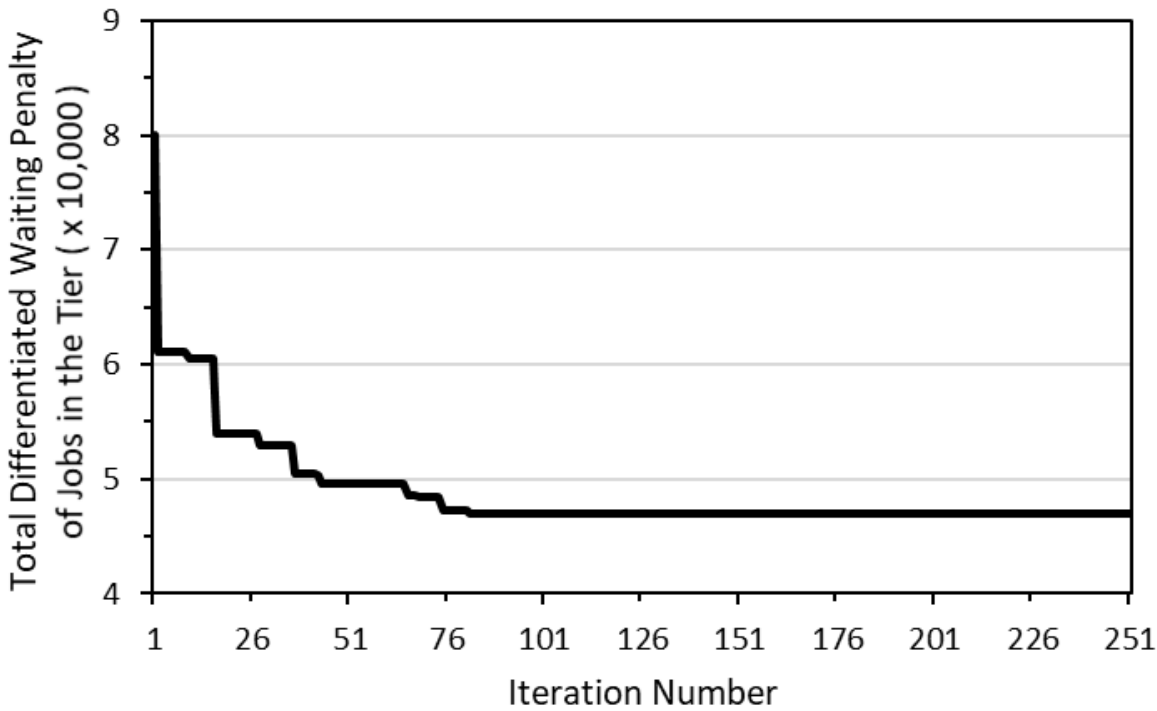}
	            \caption{Virtual Queue of 20 Jobs}
	            \label{fig:TierBased_80038}
        \end{subfigure}

        \begin{subfigure}{0.48875\textwidth}
		        \includegraphics[width=\textwidth]
                {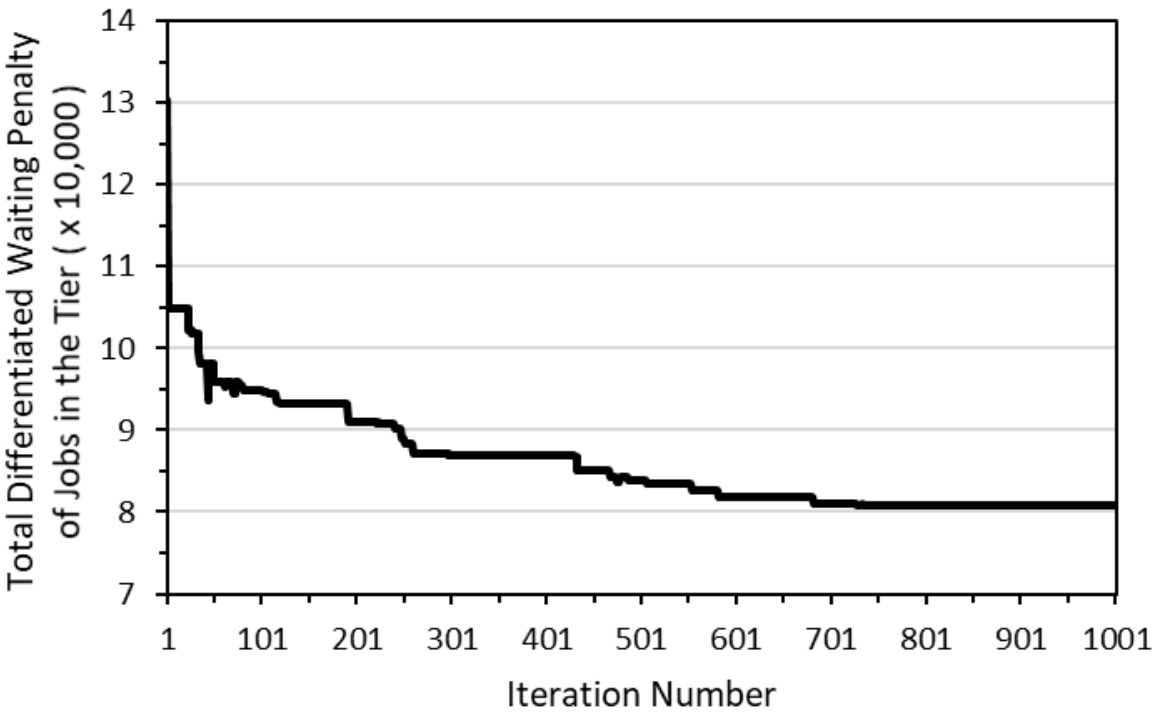}
	            \caption{Virtual Queue of 25 Jobs}
	            \label{fig:TierBased_130253}
        \end{subfigure}
        ~
        \begin{subfigure}{0.48875\textwidth}
		        \includegraphics[width=\textwidth]
                {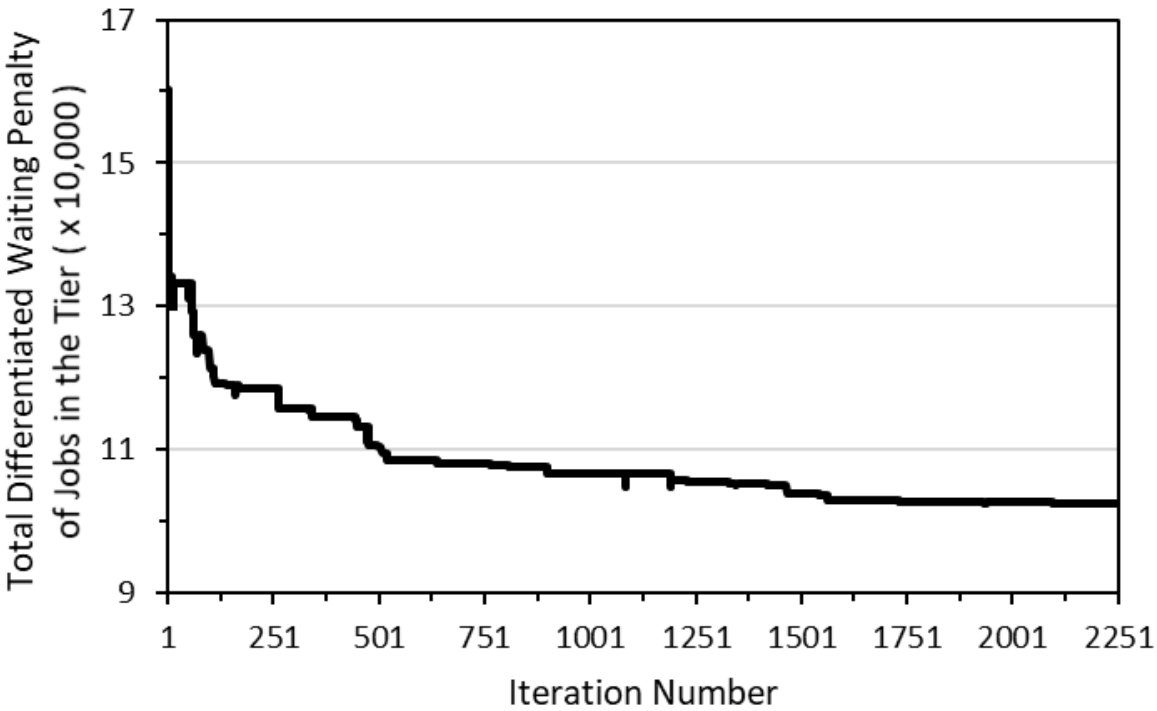}
	            \caption{Virtual Queue of 30 Jobs}
	            \label{fig:TierBased_160271}
        \end{subfigure}

        \caption{Differentiated Waiting Penalty Tier-Based Scheduling}
        \label{fig:Case3_DWPTierBased}
\end{figure*}

\begin{table*}[!ht]
\label{tab:TierBasedTable}
\caption{Differentiated Waiting Penalty Tier-Based Scheduling}
\captionsetup{justification=centering}
\begin{center}\scalebox{0.915}{
\begin{threeparttable}
\begin{tabular}{c|c|cccc|cc}
\hline
\multirow{2}{*}{}
& \multirow{2}{*}{\textbf{\begin{tabular}[c]{@{}c@{}}Virtual-Queue \\ Length\end{tabular}}}\tnote{1}
& \multicolumn{2}{c}{\textbf{Initial}\tnote{2}} & \multicolumn{2}{c|}{\textbf{Enhanced}\tnote{3}}
& \multicolumn{2}{c}{\textbf{Improvement}}  \\ \cline{3-8}
& & \textbf{Waiting}  & \textbf{Penalty} & \textbf{Waiting} & \textbf{Penalty} & \textbf{Waiting \%} & \textbf{Penalty \%} \\ \hline \hline

Figure~\ref{fig:TierBased_38203}      & 15 & 38203 & 0.318 & 21168 & 0.191 & 44.59\%   & 39.92\%       \\
Figure~\ref{fig:TierBased_80038}      & 20 & 80039 & 0.551 & 46190 & 0.370 & 42.29\%   & 32.85\%       \\ \hline

Figure~\ref{fig:TierBased_130253}     & 25 & 130253 & 0.728 & 80532  & 0.553 & 38.17\%   & 24.05\%     \\
Figure~\ref{fig:TierBased_160271}     & 30 & 160271 & 0.799 & 102137 & 0.640 & 36.27\%   & 19.88\%     \\ \hline

\end{tabular}
\begin{tablenotes}\footnotesize
\item[1] \textbf{Virtual-Queue Length} represents the total number of jobs in queues of the tier. For instance, the first entry of the table means that the 3 queues of the tier altogether are allocated 15 jobs.
\item[2] \textbf{Initial Waiting} represents the total waiting penalty of jobs in the virtual queue according to the their initial scheduling before using the tier-based genetic solution.
\item[3] \textbf{Enhanced Waiting} represents the total waiting penalty of jobs in the virtual queue according to the their final/enhanced scheduling found after using the tier-based genetic solution.
\end{tablenotes}
\end{threeparttable}}
\end{center}
\end{table*}

\subsection{Experimental Evaluation: Performance Penalty}
\label{sec:discResults3}

The optimal schedule is the one with a minimum differentiated penalty cost. The penalty cost performance of the proposed scheduling algorithm is mitigated. The effectiveness of penalty cost-driven schedules that produce optimal enhancement and consider the performance of the scheduling algorithm at the single-tier level is evaluated. The virtualized queue and segmented queue genetic scheduling are employed, as well as the service penalty function $f_{r,G}$ in Equation~\ref{sec:tierBasedVQ} is used.

\begin{figure*}[ht!]
        \centering
        \captionsetup{justification=centering}
        \begin{subfigure}{0.32\textwidth}
		        \includegraphics[width=\textwidth]
                {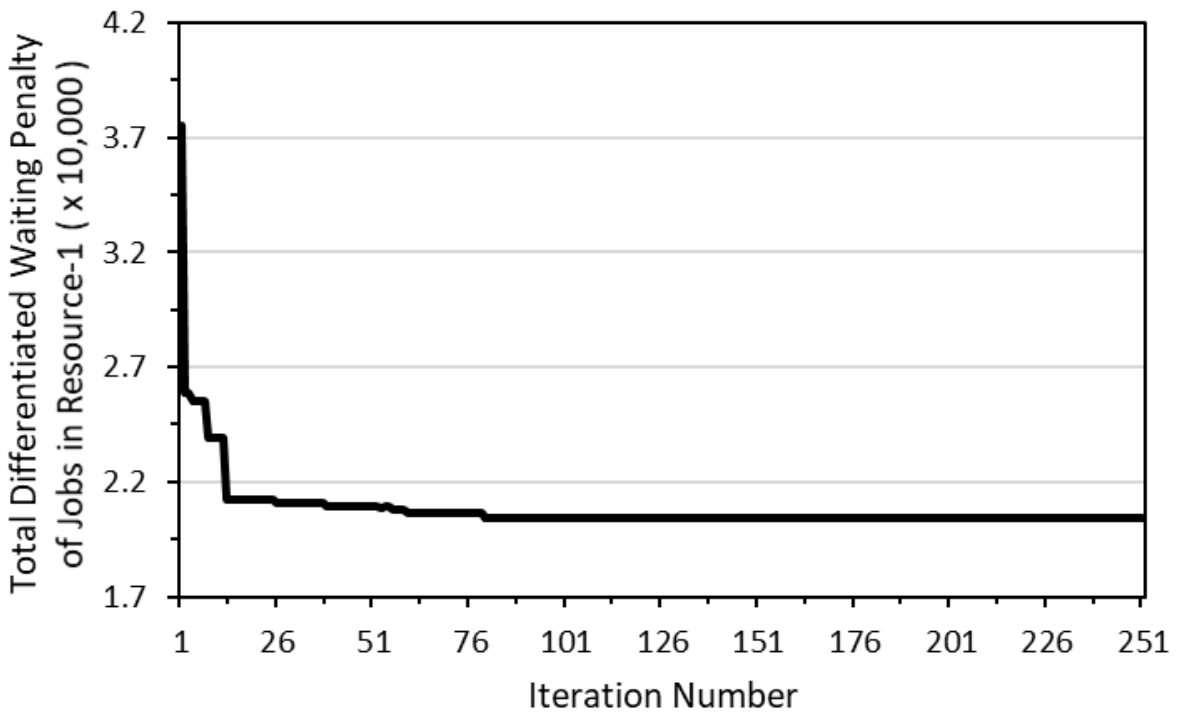}
	            \caption{Resource 1 (Queue of 8 Jobs)}
	            \label{fig:Server1_37541}
        \end{subfigure}
        ~
        \begin{subfigure}{0.32\textwidth}
		        \includegraphics[width=\textwidth]
                {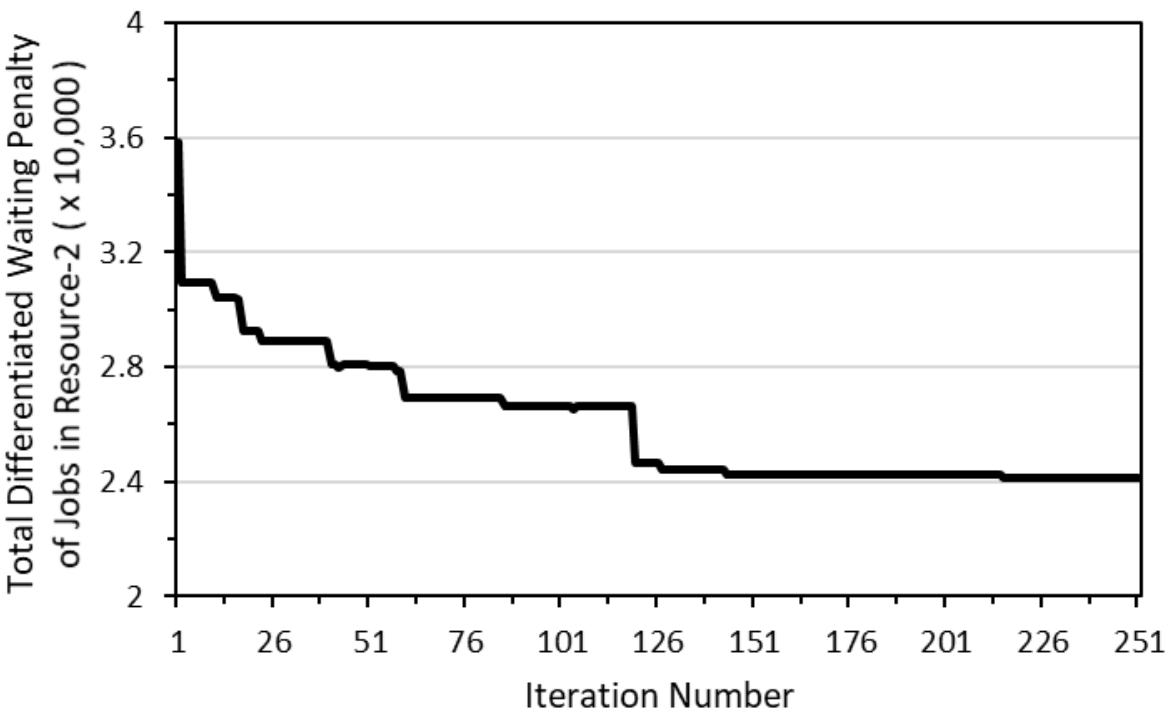}
	            \caption{Resource 2 (Queue of 10 Jobs)}
	            \label{fig:Server2_35853}
        \end{subfigure}
        ~
        \begin{subfigure}{0.32\textwidth}
                \includegraphics[width=\textwidth]
                {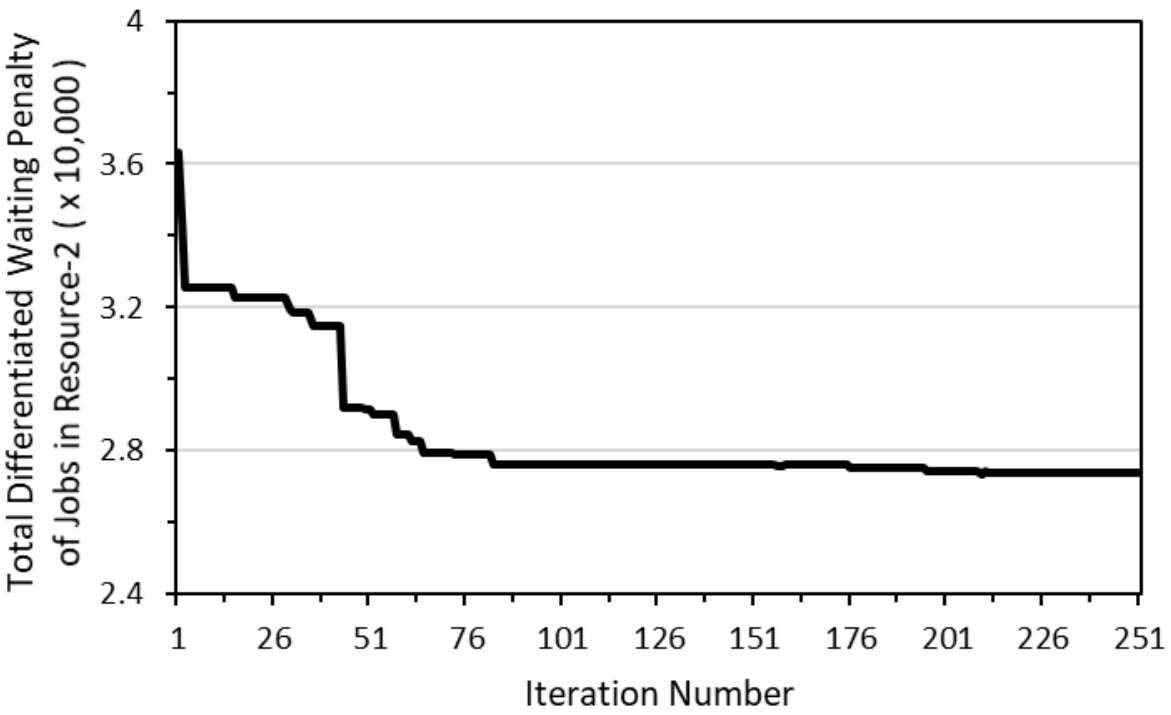}
	            \caption{Resource 3 (Queue of 13 Jobs)}
	            \label{fig:Server3_36343}
        \end{subfigure}

        \begin{subfigure}{0.32\textwidth}
		        \includegraphics[width=\textwidth]
                {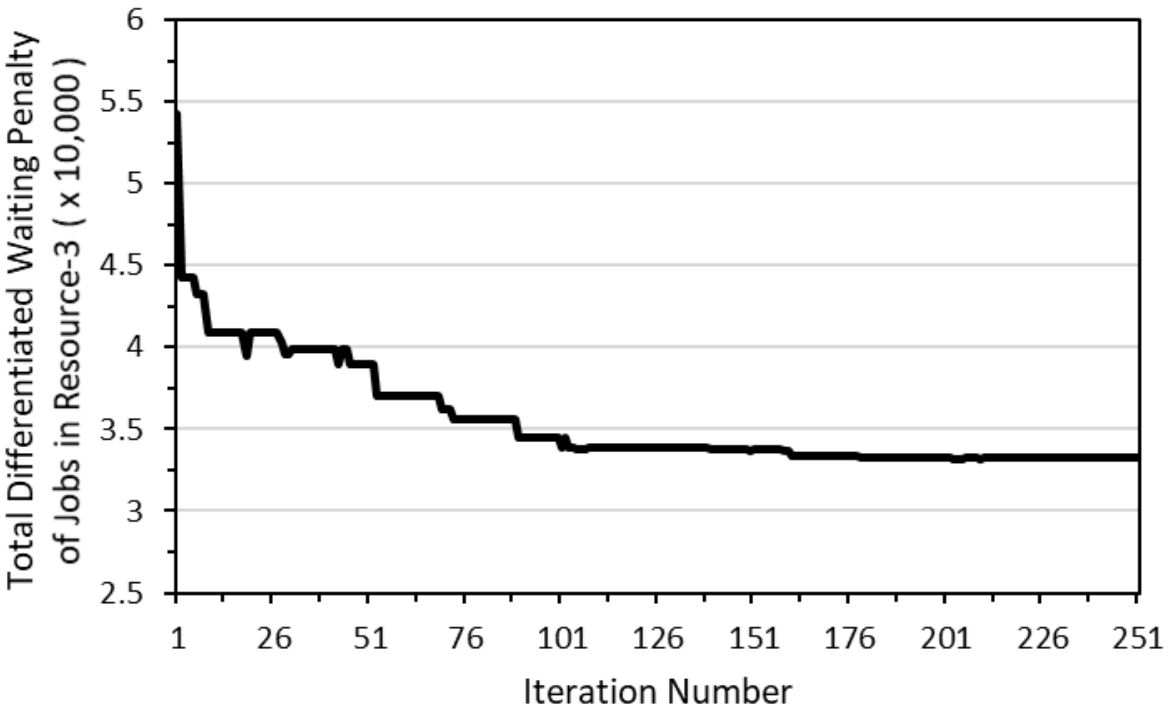}
	            \caption{Resource 1 (Queue of 12 Jobs)}
	            \label{fig:Server1_54202} 
        \end{subfigure}
        ~
        \begin{subfigure}{0.32\textwidth}
		        \includegraphics[width=\textwidth]
                {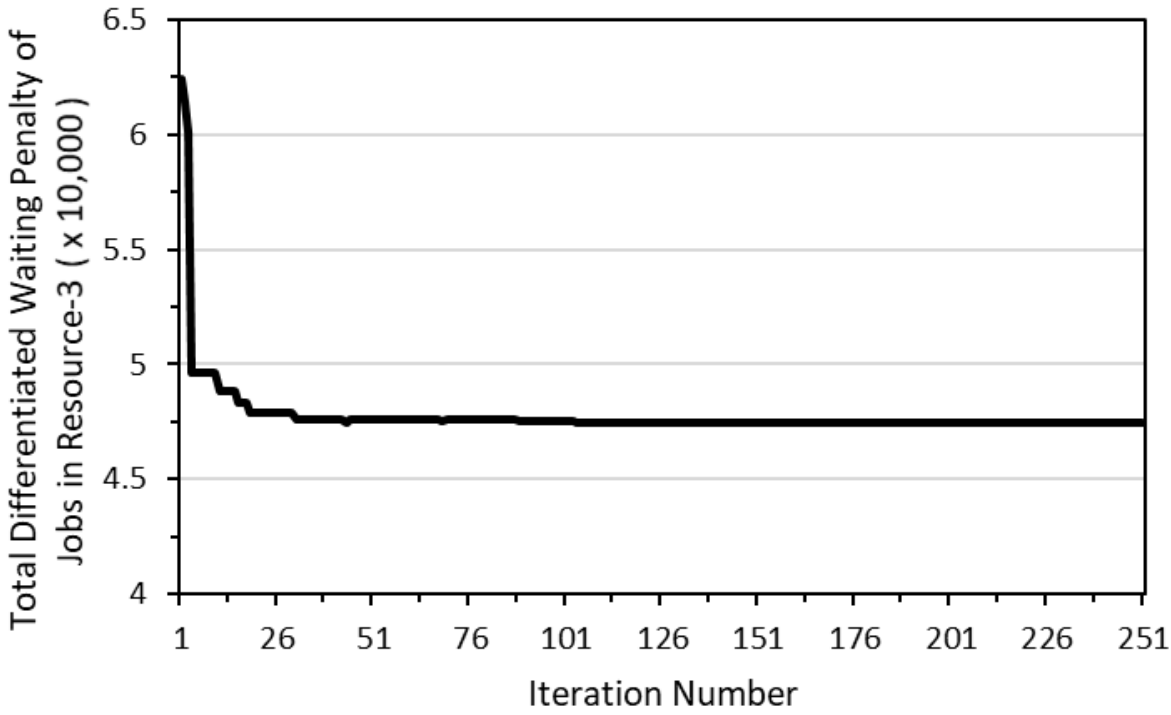}
	            \caption{Resource 2 (Queue of 8 Jobs)}
	            \label{fig:Server2_62432} 
        \end{subfigure}
        ~
        \begin{subfigure}{0.32\textwidth}
                \includegraphics[width=\textwidth]
                {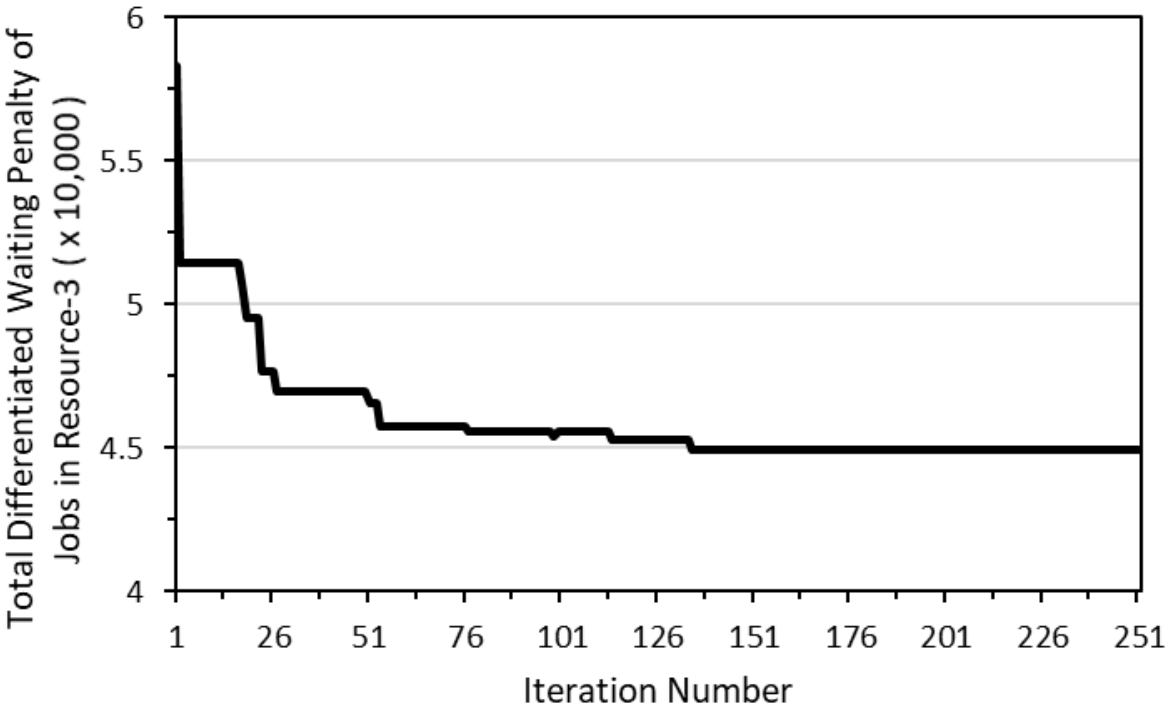}
	            \caption{Resource 3 (Queue of 9 Jobs)}
	            \label{fig:Server3_58318}
        \end{subfigure}

        \caption{Differentiated Waiting Penalty Queue-Based Scheduling}
        \label{fig:Case3_DWPQueueBased}
\end{figure*} 

\begin{table*}[!ht]
\label{tab:DWPQueueBasedTable}
\caption{Differentiated Waiting Penalty Queue-Based Scheduling}
\captionsetup{justification=centering}
\begin{center}\scalebox{0.88}{
\begin{threeparttable}
\begin{tabular}{c|c|cccc|cc}
\hline
\multirow{2}{*}{}
& \multirow{2}{*}{\textbf{\begin{tabular}[c]{@{}c@{}}Queue \\ Length\end{tabular}}}\tnote{1}
& \multicolumn{2}{c}{\textbf{Initial}\tnote{2}} & \multicolumn{2}{c|}{\textbf{Enhanced}\tnote{3}}
& \multicolumn{2}{c}{\textbf{Improvement}}  \\ \cline{3-8}
& & \textbf{Waiting}  & \textbf{Penalty} & \textbf{Waiting} & \textbf{Penalty} & \textbf{Waiting \%} & \textbf{Penalty \%} \\ \hline \hline
Resource 1 Figure~\ref{fig:Server1_37541}  & 8  & 37541 & 0.313 & 20431 & 0.185 & 45.58\% & 40.96\%   \\
Resource 2 Figure~\ref{fig:Server2_35853}  & 10 & 35853 & 0.301 & 24126 & 0.214 & 32.71\% & 28.85\%   \\
Resource 3 Figure~\ref{fig:Server3_36343}  & 13 & 36344 & 0.305 & 27162 & 0.238 & 25.26\% & 21.94\%   \\ \hline\hline
Resource 1 Figure~\ref{fig:Server1_54202}  & 12 & 54202 & 0.418 & 33130 & 0.282 & 38.88\% & 32.60\%   \\
Resource 2 Figure~\ref{fig:Server2_62432}  & 8  & 62432 & 0.464 & 47481 & 0.378 & 23.95\% & 18.60\%   \\
Resource 3 Figure~\ref{fig:Server3_58318}  & 9  & 58319 & 0.442 & 44934 & 0.362 & 22.95\% & 18.09\%   \\ \hline
\end{tabular}
\begin{tablenotes}\footnotesize
\item[1] \textbf{Queue Length} represents the number of jobs in the queue of a resource.
\item[2] \textbf{Initial Waiting} represents the total waiting penalty of jobs in the queue according to their initial scheduling before using the segmented queue genetic solution.
\item[3] \textbf{Enhanced Waiting} represents the total waiting penalty of jobs in the queue according to their final/enhanced scheduling found after using the segmented queue genetic solution.
\end{tablenotes}
\end{threeparttable}}
\end{center}
\end{table*}

The results reported in Table~1 and Figure~\ref{fig:Case3_DWPTierBased} demonstrate the effectiveness of the differentiated penalty-based scheduling in reducing total service penalty cost, at the virtualized queue level. For instance, the penalty of the initial scheduling shown in Figure~\ref{fig:TierBased_38203} has a cost of $38,\!203$ time units. The differentiated penalty scheduling algorithm produces schedules that reduce this cost by $44.59\%$, to $21,\!168$ units. Consequently, the SLA penalty payable by the cloud service provider has also been improved by $39.92\%$, a reduction from $0.381$ for the initial scheduling to $0.191$ for the enhanced penalty-based scheduling.

In addition, the differentiated penalty-based scheduling demonstrates its effectiveness in optimizing financial performance by formulating cost-optimal schedules at the individual-queue level, as shown in Table~2 and Figure~\ref{fig:Case3_DWPQueueBased}. For example, resource-3 (presented in Figure~\ref{fig:Server3_36343}) demonstrates the efficacy of the penalty-based scheduling in improving the penalty cost of the job schedule by $25\%$, a reduction in cost from $36,\!344$ to $27,\!126$ time units. As a result, the performance of the differentiated penalty cost of the queue-state is optimized by $21.94\%$, reduced from $0.305$ due to the initial scheduling order to reach $0.238$ due to the improved differentiated penalty-based schedule. 

Thus, the virtualized queue and segmented queue genetic solutions have efficiently explored a large solution search space using a small number of genetic iterations to achieve the enhancements. Figure~\ref{fig:TierBased_130253} shows that the virtualized queue required a total of only $1,\!000$ genetic iterations to efficiently seek an optimal schedule of jobs in tier $T\!_1$, each iteration employs 10 chromosomes to evolve the optimal schedule. As such, $\nolinebreak{10\!\times\!10^3}$ scheduling orders are constructed and genetically manipulated throughout the search space, as opposed to $25!$ (approximately $\nolinebreak{1.55\!\times\!10^{25}}$) scheduling orders if a brute-force search strategy is employed to seek the optimal scheduling of jobs. Similar observations are in order with respect to the results reported on the segmented queue genetic solution.
\begin{table*}[!h]
\label{tab:comparison4}
\centering
\caption{Total Differentiated Waiting Penalty}
\captionsetup{justification=centering}
\scalebox{0.99}{
\begin{tabular}{cccccc}
\hline

\textbf{\begin{tabular}[c]{@{}c@{}}Differentiated Penalty\\ Virtualized Queue\end{tabular}} & \textbf{\begin{tabular}[c]{@{}c@{}}Differentiated Penalty\\ Segmented Queue\end{tabular}} & \textbf{\begin{tabular}[c]{@{}c@{}}Virtualized\\ Queue\end{tabular}} &
\textbf{\begin{tabular}[c]{@{}c@{}}Segmented\\ Queue\end{tabular}} &
\textbf{WLC} &
\textbf{WRR} \\ \hline

2423344  &  2709716  &  2976390  &  3004961  &  3652770  &  3899232  \\ \hline

\end{tabular}}
\end{table*}
\begin{figure*}[!th]
\centering \includegraphics[width=\textwidth]{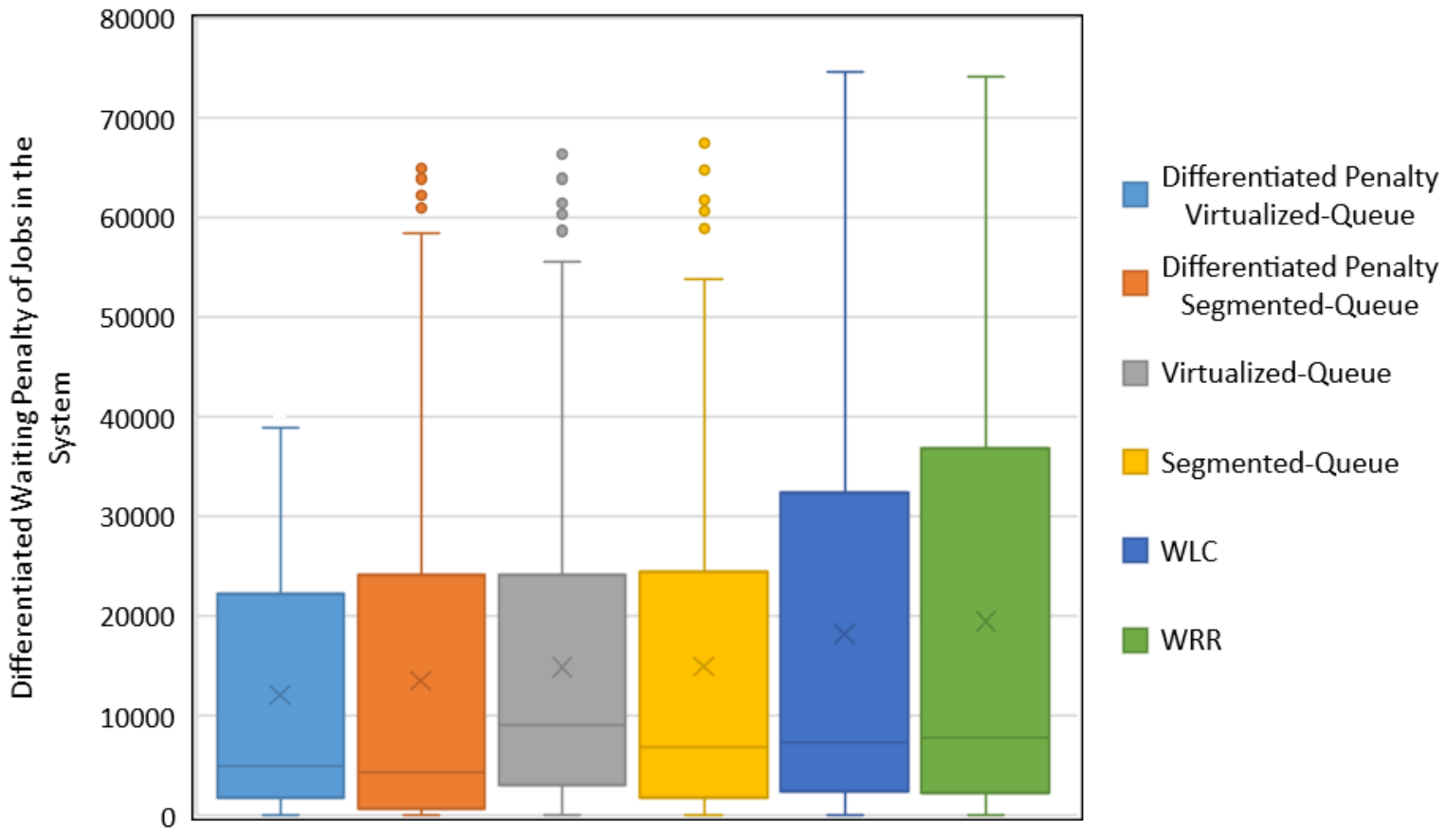}
	 \caption{Maximum Differentiated Waiting Penalty Performance Comparison}
     \label{fig:Paper3_Comparison}
\end{figure*}

To contrast the financial performance of the scheduling strategies, Table~3 and Figure~\ref{fig:Paper3_Comparison} evaluate the differentiated service penalty cost. The WLC and WRR entail a cost of $\nolinebreak{3.65\!\times\!10^6}$ and $\nolinebreak{3.9\!\times\!10^6}$ time units, respectively. However, the virtualized queue and segmented queue scheduling approaches (without the service cost $\psi_i$) show superior performance compared with WLC and WRR, yet show inferior performance in improving the service penalty cost compared with the differentiated penalty-based scheduling approaches.

In fact, the differentiated penalty-based virtualized and segmented queue scheduling approaches produce schedules that improve service penalty cost. The differentiated penalty-based scheduling of the segmented queue genetic approach reduces the service penalty to a cost of $\nolinebreak{2.7\!\times\!10^6}$ time units, demonstrating a superior performance compared with WLC and WRR. In contrast, the differentiated penalty-based scheduling of the virtualized queue genetic approach optimizes financial performance by reducing service penalty cost to $\nolinebreak{2.4\!\times\!10^6}$, demonstrating the best financial performance compared with the other scheduling strategies.

Overall, single-tier-driven differentiated penalty scheduling produces schedules that enhance financial performance. The virtualized queue and segmented queue genetic approaches employed in the scheduling process demonstrate their effectiveness in efficiently facilitating the search for financially performance-optimal schedules at the tier level and individual queue level of the tier, respectively.

\subsection{Evaluation of Differentiated Scheduling: Multi-Tier Considerations}
\label{sec:discResults4}

This is concerned with formulating performance-optimal schedules that produce a minimum differentiated SLA penalty at the multi-tier level. The experiments are conducted using the system virtualized queue and segmented queue genetic scheduling, explained in section~\ref{sec:MultitierBasedVQ}. The QoS penalty function $f_{r,G}$ of the multi-tier genetic scheduling in Equation~\ref{equ:fitness4} is used. Thus, the penalty function evaluates the effectiveness of schedules to reach an optimal financial performance by minimizing the differentiated multi-tier SLA penalty.

\begin{figure*}[!t]
        \centering
        \captionsetup{justification=centering}
        \begin{subfigure}{0.85\textwidth}
		        \includegraphics[width=\textwidth]
                {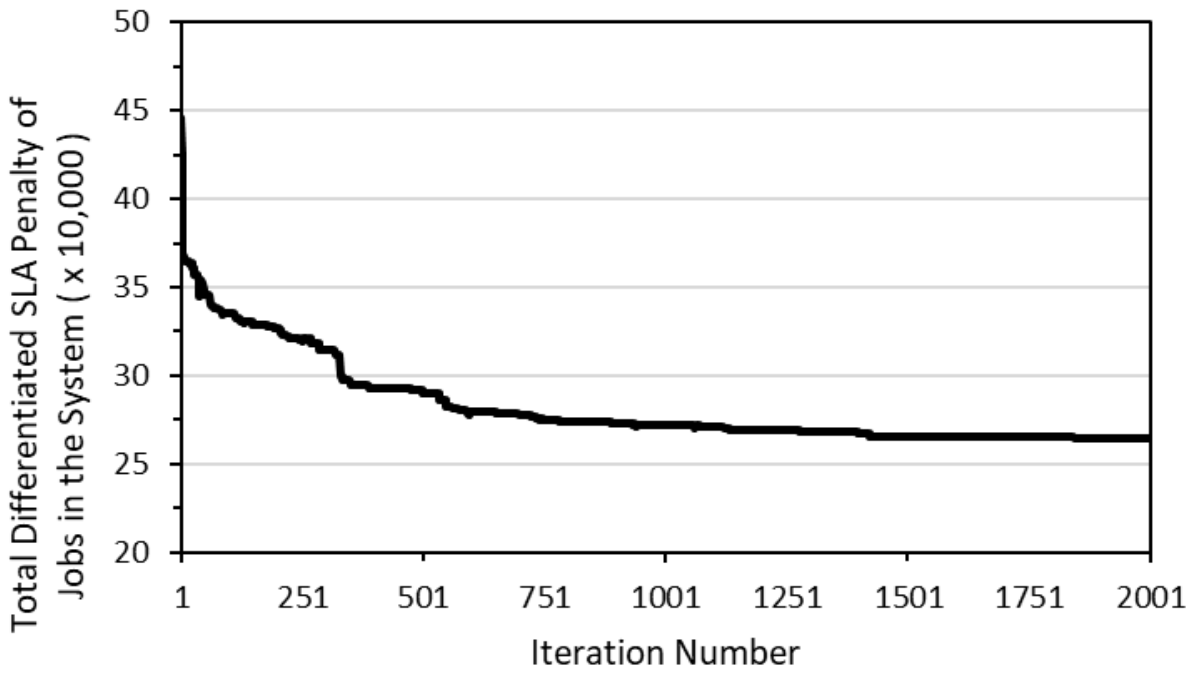}
	            \caption{System-Level (Total of 69 Jobs)}
	            \label{fig:SYSTEM_446182}
        \end{subfigure}

        \begin{subfigure}{0.48875\textwidth}
		        \includegraphics[width=\textwidth]
                {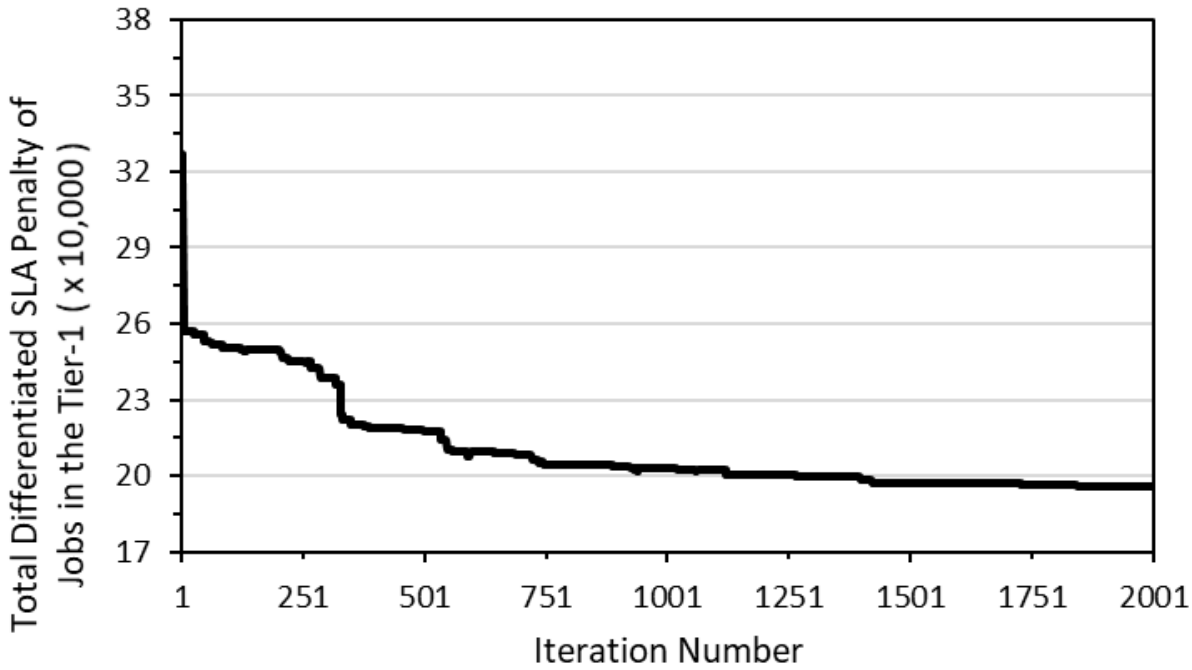}
	            \caption{Tier-1 (40 Jobs)}
	            \label{fig:Tier1SYSTEM_327232}
        \end{subfigure}
        ~
        \begin{subfigure}{0.48875\textwidth}
                \includegraphics[width=\textwidth]
                {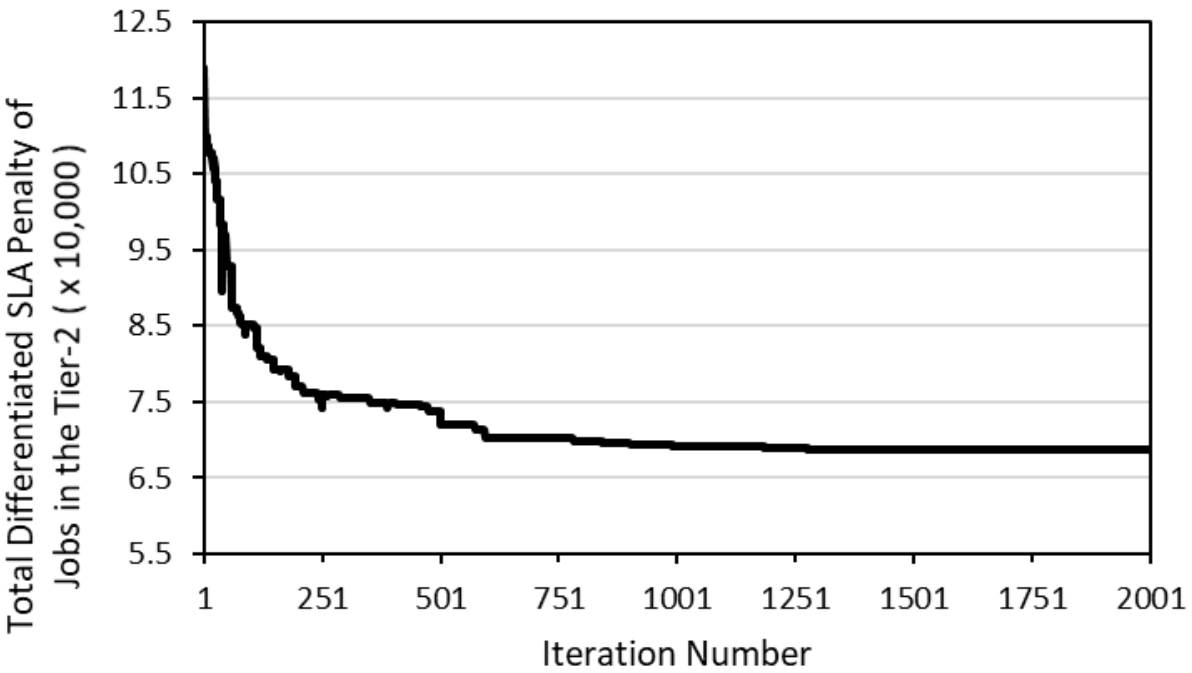}
	            \caption{Tier-2 (29 Jobs)}
	            \label{fig:Tier2SYSTEM_118950}
        \end{subfigure}
        \caption{Differentiated SLA Penalty Multi-Tier $\omega\!\mathcal{A\!L}_i$ Based System Virtualized Queue Scheduling}
        \label{fig:Case4_DslaPtierSYSTEM}
\end{figure*}

\begin{table*}[!ht]
\label{tab:Tier_DslaPSystemLevel}
\captionsetup{justification=centering}
\caption{Differentiated SLA Penalty Multi-Tier $\omega\!\mathcal{A\!L}_i$ Based System Virtualized Queue Scheduling}
\begin{center}\scalebox{0.82}{
\begin{threeparttable}
\begin{tabular}{c|c|cccc|cc}
\hline
\multirow{2}{*}{}
& \multirow{2}{*}{\textbf{\begin{tabular}[c]{@{}c@{}}Number \\of Jobs\end{tabular}}}\tnote{1}
& \multicolumn{2}{c}{\textbf{Initial}\tnote{2}} & \multicolumn{2}{c|}{\textbf{Enhanced}\tnote{3}}
& \multicolumn{2}{c}{\textbf{Improvement}}  \\ \cline{3-8}
& & \textbf{Violation}  & \textbf{Penalty} & \textbf{Violation} & \textbf{Penalty} & \textbf{Violation \%} & \textbf{Penalty \%} \\ \hline \hline
System-Level, Figure~\ref{fig:SYSTEM_446182}
& 69 & 446183 & 1.66 & 262387 & 1.35 & 41.19\% & 18.38\%   \\ \hline
Tier-1, Figure~\ref{fig:Tier1SYSTEM_327232}
& 40 & 327232 & 0.96 & 193614 & 0.86 & 40.83\% & 11.05\%   \\
Tier-2, Figure~\ref{fig:Tier2SYSTEM_118950}
& 29 & 118951 & 0.70 & 68773  & 0.50 & 42.18\% & 28.51\%   \\ \hline
\end{tabular}
\begin{tablenotes}\scriptsize
\item[1] \textbf{Number of Jobs} represents the total number of jobs in queues of the tier/environment. The multi-tier environment contains 69 jobs in total. The 3 queues of tier-1 and tier-2 are allocated 40 and 29 jobs, respectively.
\item[2] \textbf{Initial Violation} represents the total SLA violation time of jobs according to their initial scheduling before using the system virtualized queue genetic solution.
\item[3] \textbf{Enhanced Violation} represents the total SLA violation time of jobs according to their final/enhanced scheduling found after using the system virtualized queue genetic solution.
\end{tablenotes}
\end{threeparttable}}
\end{center}
\end{table*} 

\begin{figure*}[!ht]
        \centering
        \captionsetup{justification=centering}
        \begin{subfigure}{0.6\textwidth}
		        \includegraphics[width=\textwidth]
                {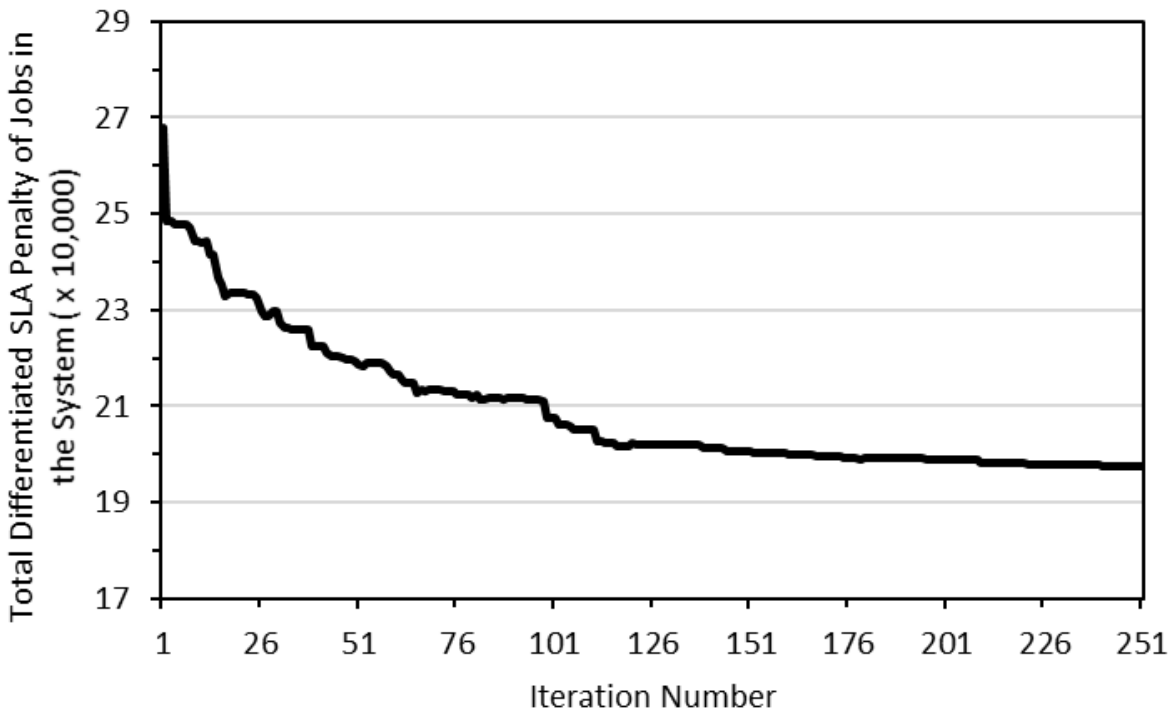}
	            \caption{System-Level (Total of 75 Jobs)}
	            \label{fig:SYSTEM_267775}
        \end{subfigure}

        \begin{subfigure}{0.315\textwidth}
		        \includegraphics[width=\textwidth]
                {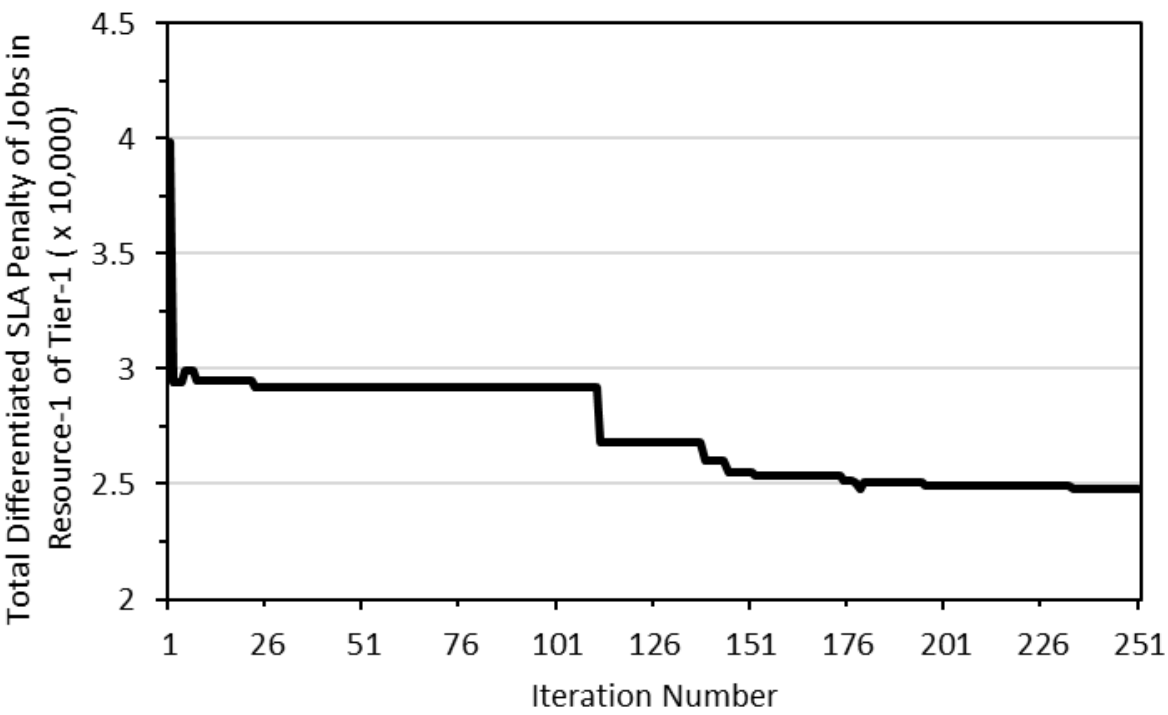}
	            \caption{Resource 1 of Tier 1\\~~~(Queue of 9 Jobs)}
	            \label{fig:Server1Tier1SYSTEM_39837} 
        \end{subfigure}
        ~
        \begin{subfigure}{0.315\textwidth}
		        \includegraphics[width=\textwidth]
                {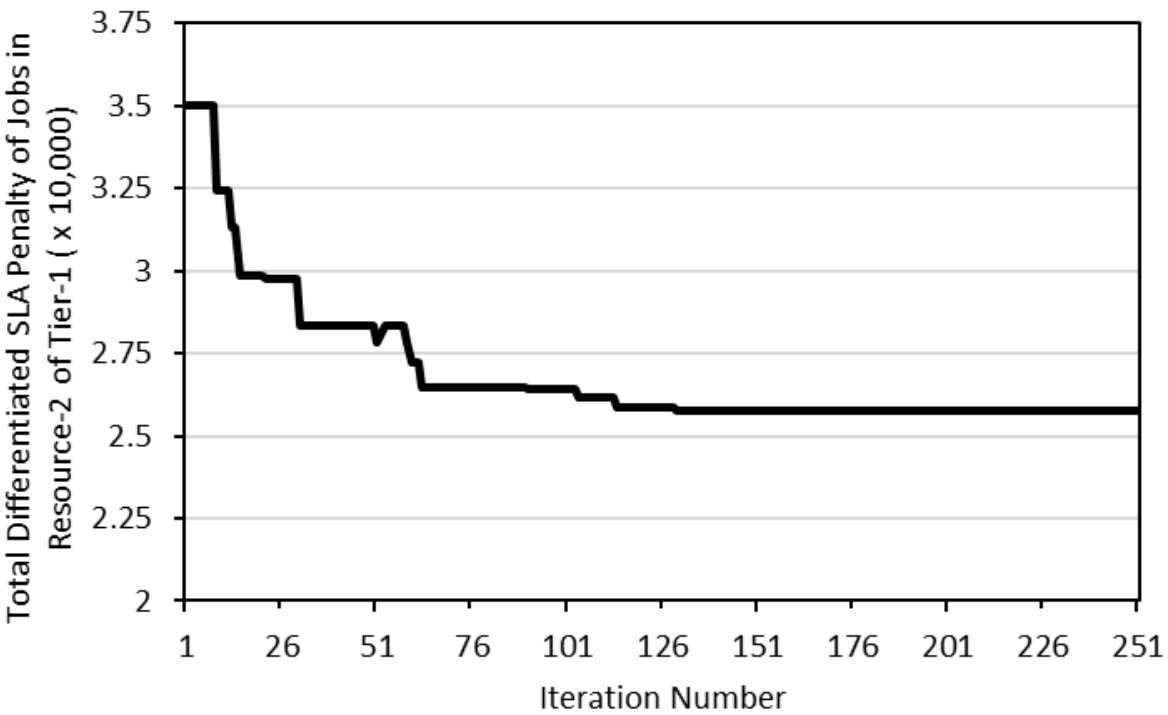}
	            \caption{Resource 2 of Tier 1\\~~~(Queue of 13 Jobs)}
	            \label{fig:Server2Tier1SYSTEM_34987} 
        \end{subfigure}
        ~
        \begin{subfigure}{0.315\textwidth}
                \includegraphics[width=\textwidth]
                {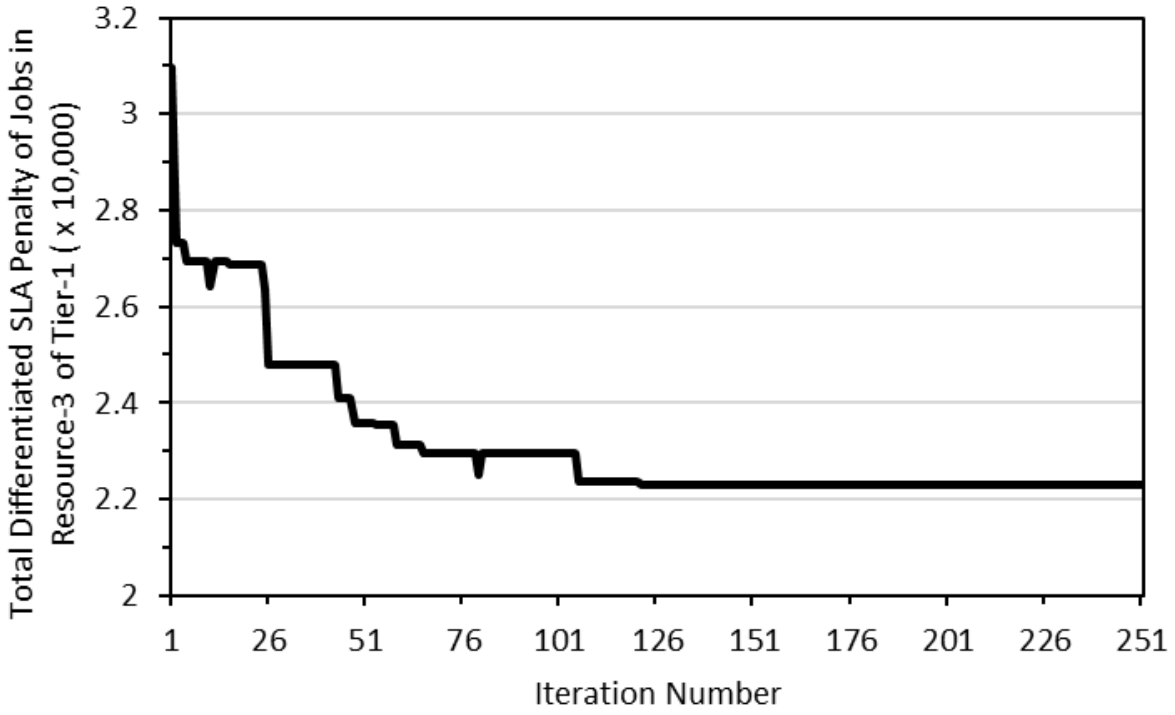}
	            \caption{Resource 3 of Tier 1\\~~~(Queue of 10 Jobs)}
	            \label{fig:Server3Tier1SYSTEM_30975}
        \end{subfigure}

        \begin{subfigure}{0.315\textwidth}
		        \includegraphics[width=\textwidth]
                {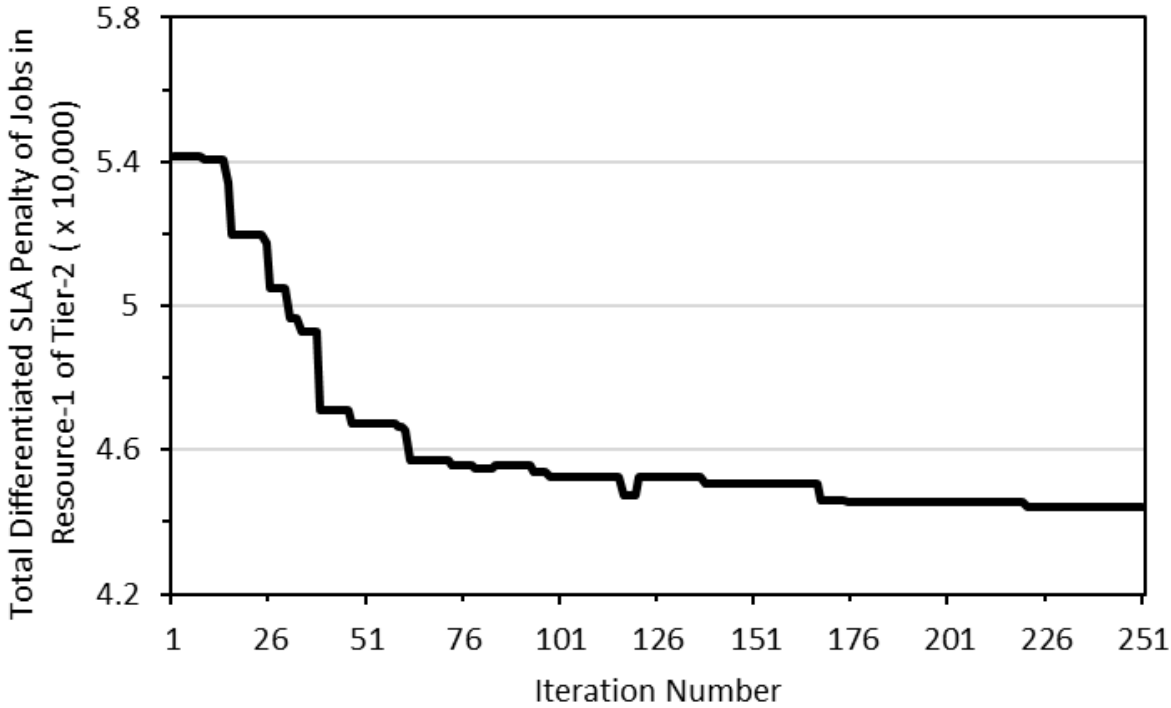}
	            \caption{Resource 1 of Tier 2\\~~~(Queue of 13 Jobs)}
	            \label{fig:Server1Tier2SYSTEM_54131} 
        \end{subfigure}
        ~
        \begin{subfigure}{0.315\textwidth}
		        \includegraphics[width=\textwidth]
                {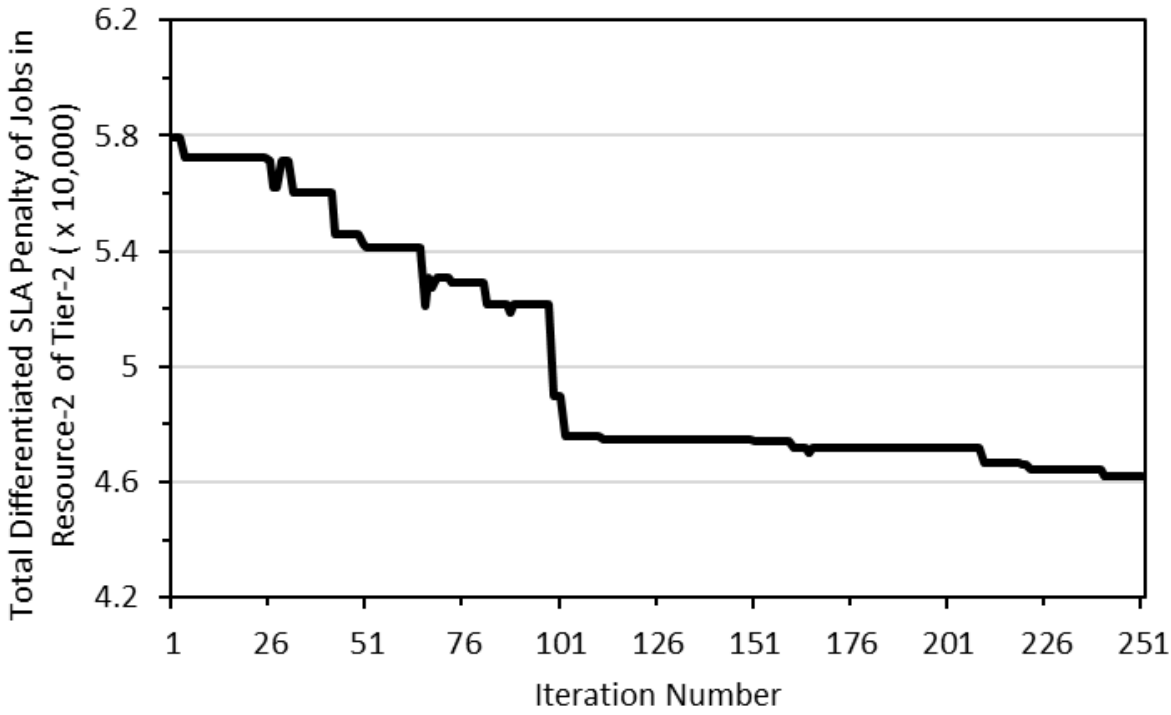}
	            \caption{Resource 2 of Tier 2\\~~~(Queue of 16 Jobs)}
	            \label{fig:Server2Tier2SYSTEM_57944} 
        \end{subfigure}
        ~
        \begin{subfigure}{0.315\textwidth}
                \includegraphics[width=\textwidth]
                {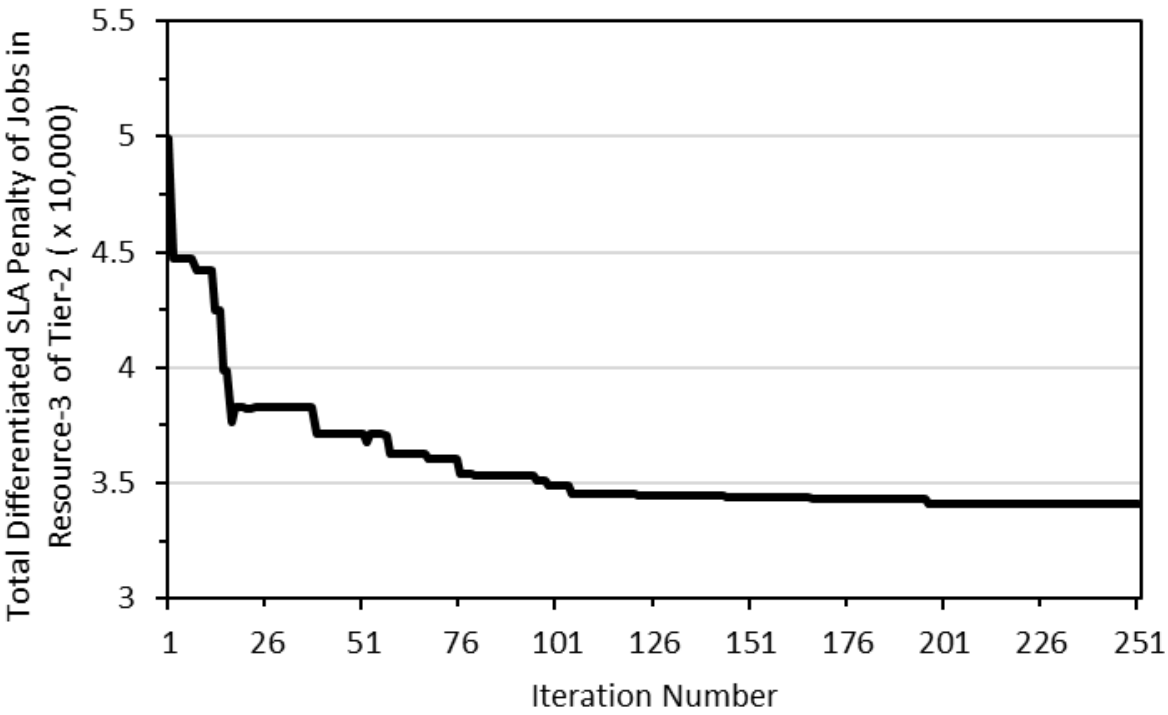}
	            \caption{Resource 3 of Tier 2\\~~~(Queue of 14 Jobs)}
	            \label{fig:Server3Tier2SYSTEM_49898}
        \end{subfigure}

        \caption{Differentiated SLA Penalty Multi-Tier $\omega\!\mathcal{A\!L}_i$ Based Segmented Queue Scheduling}
        \label{fig:Case4_DslaPqueueSYSTEM}
\end{figure*} 

\begin{table*}[!ht]
\label{tab:Queue_SystemLevel}
\caption{Differentiated SLA Penalty Multi-Tier $\omega\!\mathcal{A\!L}_i$ Based Segmented Queue Scheduling}
\captionsetup{justification=centering}
\begin{center}\scalebox{0.8}{
\begin{threeparttable}
\begin{tabular}{c|c|cccc|cc}
\hline
\multirow{2}{*}{} & \multirow{2}{*}{\textbf{\begin{tabular}[c]{@{}c@{}}Number \\of Jobs\end{tabular}}}
& \multicolumn{2}{c}{\textbf{Initial}\tnote{1}}
& \multicolumn{2}{c|}{\textbf{Enhanced}\tnote{2}}
& \multicolumn{2}{c}{\textbf{Improvement}}  \\ \cline{3-8}
& & \textbf{Violation}  & \textbf{Penalty} & \textbf{Violation} & \textbf{Penalty} & \textbf{Violation \%} & \textbf{Penalty \%} \\ \hline \hline

System-Level, Figure~\ref{fig:SYSTEM_267775}
& 75 & 267775 & 2.14 & 196484 & 1.66 & 26.62\% & 22.60\%   \\ \hline

Resource-1 Tier-1, Figure~\ref{fig:Server1Tier1SYSTEM_39837}
& 9  & 39837 & 0.33 & 24775 & 0.22 & 37.81\% & 33.22\%  \\
Resource-2 Tier-1, Figure~\ref{fig:Server2Tier1SYSTEM_34987}
& 13 & 34988 & 0.30 & 25724 & 0.23 & 26.48\% & 23.17\%  \\
Resource-3 Tier-1, Figure~\ref{fig:Server3Tier1SYSTEM_30975}
& 10 & 30976 & 0.27 & 22281 & 0.20 & 28.07\% & 25.02\%  \\ \hline

Resource-1 Tier-2, Figure~\ref{fig:Server1Tier2SYSTEM_54131}
& 13 & 54131 & 0.42 & 44182 & 0.36 & 18.38\% & 14.56\%   \\
Resource-2 Tier-2, Figure~\ref{fig:Server2Tier2SYSTEM_57944}
& 16 & 57945 & 0.44 & 45633 & 0.37 & 21.25\% & 16.69\%   \\
Resource-3 Tier-2, Figure~\ref{fig:Server3Tier2SYSTEM_49898}
& 14 & 49899 & 0.39 & 33890 & 0.29 & 32.08\% & 26.83\%   \\ \hline

\end{tabular}
\begin{tablenotes}\scriptsize
\item[1] \textbf{Initial Violation} represents the total SLA violation time of jobs according to their initial scheduling before using the segmented queue genetic solution.
\item[2] \textbf{Enhanced Violation} represents the total SLA violation time of jobs according to their final/enhanced scheduling found after using the segmented queue genetic solution.
\end{tablenotes}
\end{threeparttable}}
\end{center}
\end{table*}

\begin{figure*}[!ht]
        \centering
        \captionsetup{justification=centering}
        \begin{subfigure}{0.85\textwidth}
		        \includegraphics[width=\textwidth]
                {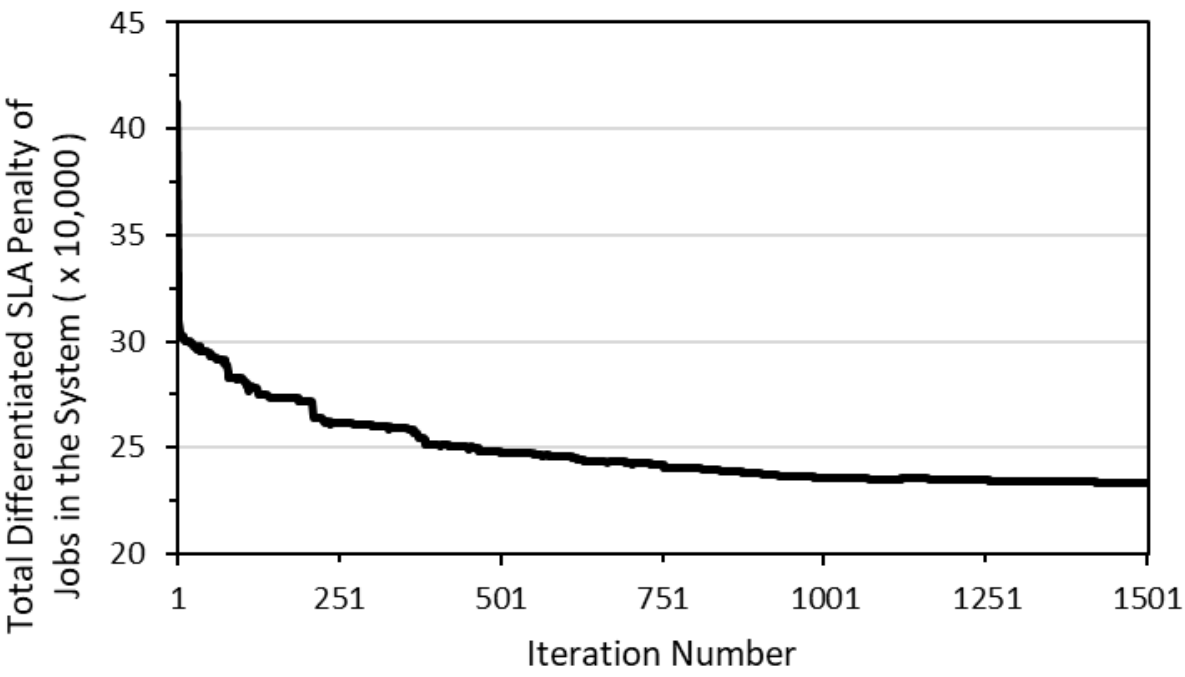}
	            \caption{System-Level (Total of 66 Jobs)}
	            \label{fig:SYSTEMproportional_412442} 
        \end{subfigure}

        \begin{subfigure}{0.48875\textwidth}
		        \includegraphics[width=\textwidth]
                {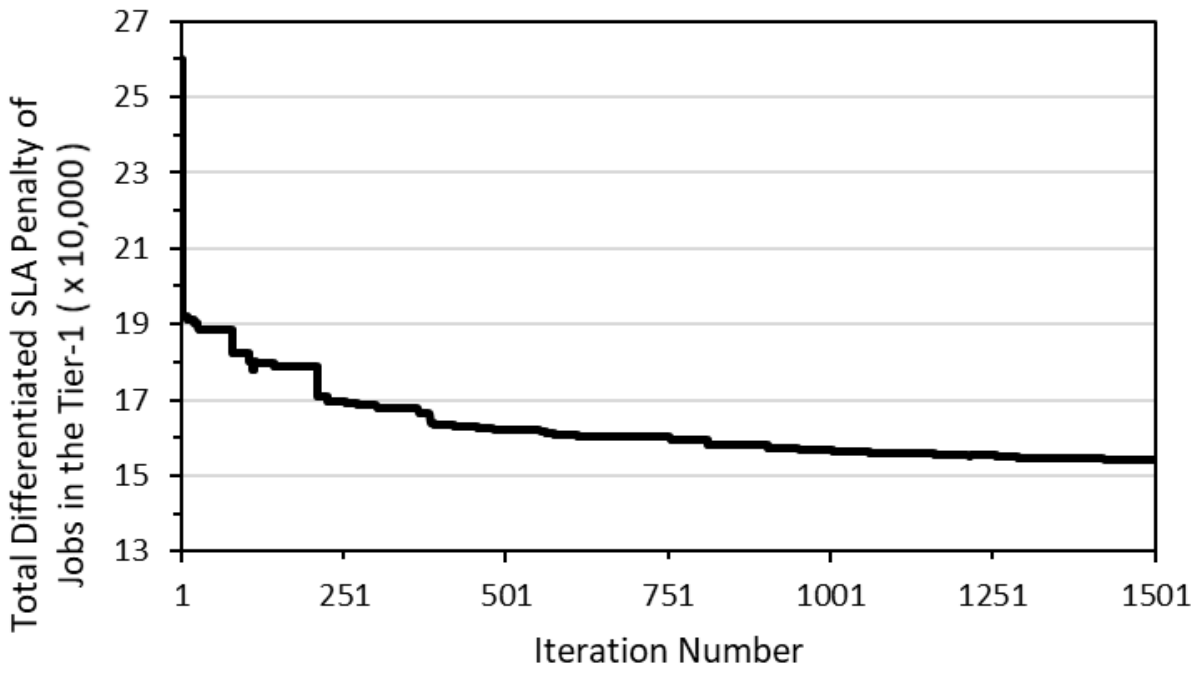}
	            \caption{Tier-1 (35 Jobs)}
	            \label{fig:Tier1SYSTEMproportional_259880} 
        \end{subfigure}
        ~
        \begin{subfigure}{0.48875\textwidth}
                \includegraphics[width=\textwidth]
                {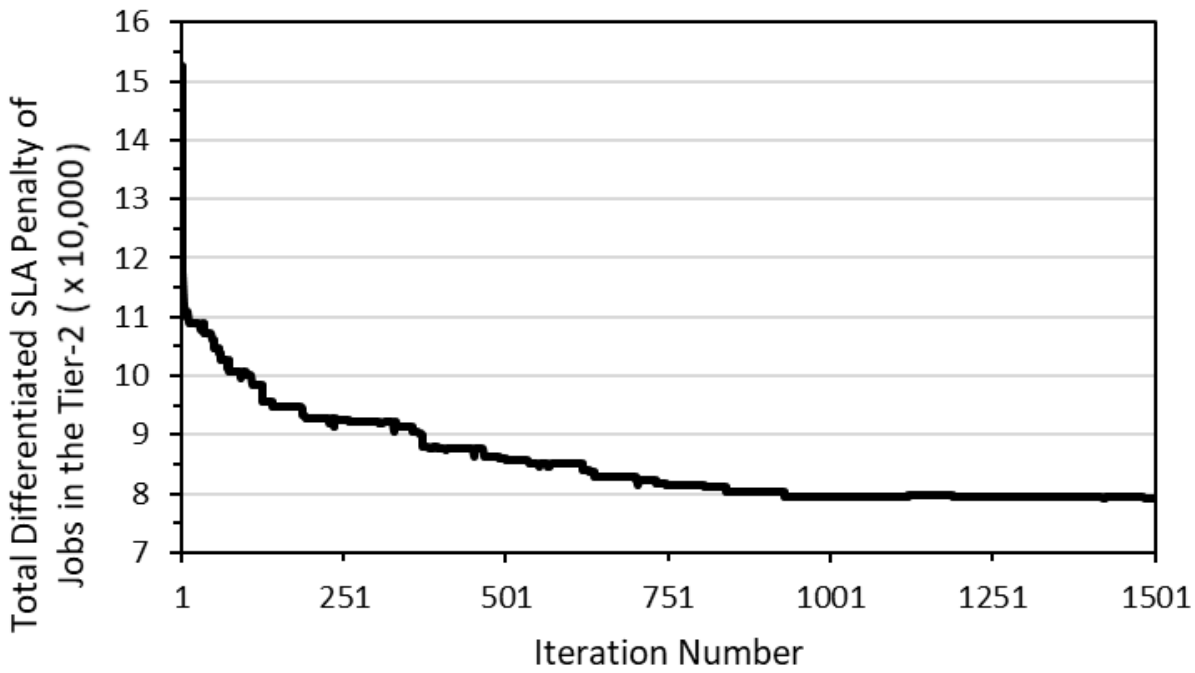}
	            \caption{Tier-2 (31 Jobs)}
	            \label{fig:Tier2SYSTEMproportional_152562}
        \end{subfigure}

        \caption{Differentiated SLA Penalty $\omega\!\mathcal{P\!T}\!_{i,j}$ Based System Virtualized Queue Scheduling}
        \label{fig:Case4_TierDslaPSYSTEMproportional}
\end{figure*} 

\begin{table*}[!ht]
\label{tab:Tier12}
\caption{Differentiated SLA Penalty $\omega\!\mathcal{P\!T}\!_{i,j}$ Based System Virtualized Queue Scheduling}
\captionsetup{justification=centering}
\begin{center}\scalebox{0.7}{
\begin{threeparttable}
\begin{tabular}{c|c|cccc|cc}
\hline
\multirow{2}{*}{}
& \multirow{2}{*}{\textbf{\begin{tabular}[c]{@{}c@{}}Number \\of Jobs\end{tabular}}}\tnote{1}
& \multicolumn{2}{c}{\textbf{Initial}\tnote{2}} & \multicolumn{2}{c|}{\textbf{Enhanced}\tnote{3}}
& \multicolumn{2}{c}{\textbf{Improvement}}  \\ \cline{3-8}
& & \textbf{Violation}  & \textbf{Penalty} & \textbf{Violation} & \textbf{Penalty} & \textbf{Violation \%} & \textbf{Penalty \%} \\ \hline \hline

System-Level, Figure~\ref{fig:SYSTEMproportional_412442}
& 66 & 412442 & 1.71 & 232573 & 1.33 & 43.61\%  & 22.05\%   \\ \hline
Tier-1, Figure~\ref{fig:Tier1SYSTEMproportional_259880}
& 35 & 259880 & 0.93 & 153300 & 0.78 & 41.01\%  & 15.29\%   \\
Tier-2, Figure~\ref{fig:Tier2SYSTEMproportional_152562}
& 31 & 152562 & 0.78 & 79273  & 0.55 & 48.04\%  & 30.05\%   \\ \hline

\end{tabular}
\begin{tablenotes}\scriptsize
\item[1] \textbf{Number of Jobs} represents the total number of jobs in queues of the tier/environment. The multi-tier environment is allocated 66 jobs in total. The 3 queues of tier-1 and tier-2 are allocated 35 and 31 jobs, respectively.
\item[2] \textbf{Initial Violation} represents the total SLA violation time of jobs according to their initial scheduling before using the system virtualized queue genetic solution.
\item[3] \textbf{Enhanced Violation} represents the total SLA violation time of jobs according to their final/enhanced scheduling found after using the system virtualized queue genetic solution.
\end{tablenotes}
\end{threeparttable}}
\end{center}
\end{table*}

The results shown in Table~4 and Figure~\ref{fig:Case4_DslaPtierSYSTEM} represent a system-state of a multi-tier environment that is allocated $69$ jobs; $40$ jobs are allocated to tier $T\!_1$ and $29$ jobs are allocated to tier $T\!_2$. The differentiated multi-tier penalty $\omega\!\mathcal{A\!L}_i$ based scheduling of the system virtualized queue genetic approach has gradually reduced the SLA penalty cost. The differentiated $\omega\!\mathcal{A\!L}_i$ based scheduling genetic evaluation function in Equation~\ref{equ:fitness4} is employed. The financial performance of the system-state is optimized by $41.19\%$, through formulating an enhanced cost-optimal schedule that reduces the SLA penalty from a cost of $446,\!183$ time units for the initial schedule to a cost of $262,\!387$ time units for the improved schedule computed at the multi-tier level. As such, the differentiated SLA penalty cost payable by the cloud service provider has been improved by $18.38\%$, a reduction in the penalty from $1.66$ for the initial schedule to $1.35$ for the improved cost-optimal schedule of the system-state.

Similarly, the differentiated multi-tier penalty $\omega\!\mathcal{A\!L}_i$ based scheduling of the segmented queue genetic approach shows an improved financial performance on the system-state. Cost-optimal schedules are formulated in each individual queue to efficiently reduce the differentiated SLA penalty cost at the multi-tier level, as shown in Table~5 and Figure~\ref{fig:Case4_DslaPqueueSYSTEM}. In a multi-tier environment allocated $75$ jobs, the differentiated SLA penalty improves by $22.6\%$ at the multi-tier level. The SLA penalty cost of the system-state has been reduced from $2.14$ for the initial schedule to reach $1.66$ for the cost-optimal schedule.

\begin{figure*}[!ht]
        \centering
        \captionsetup{justification=centering}
        \begin{subfigure}{0.6\textwidth}
		        \includegraphics[width=\textwidth]
                {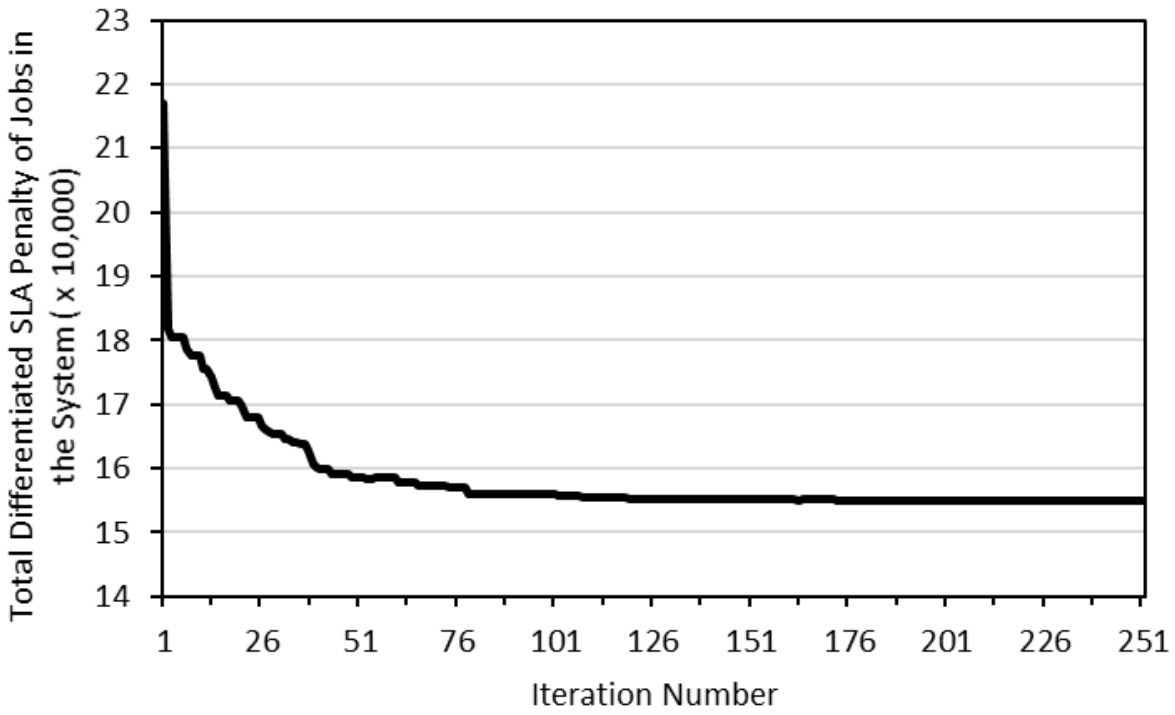}
	            \caption{System-Level (Total of 57 Jobs)}
	            \label{fig:SYSTEMproportional_216897}
        \end{subfigure}

        \begin{subfigure}{0.315\textwidth}
		        \includegraphics[width=\textwidth]
                {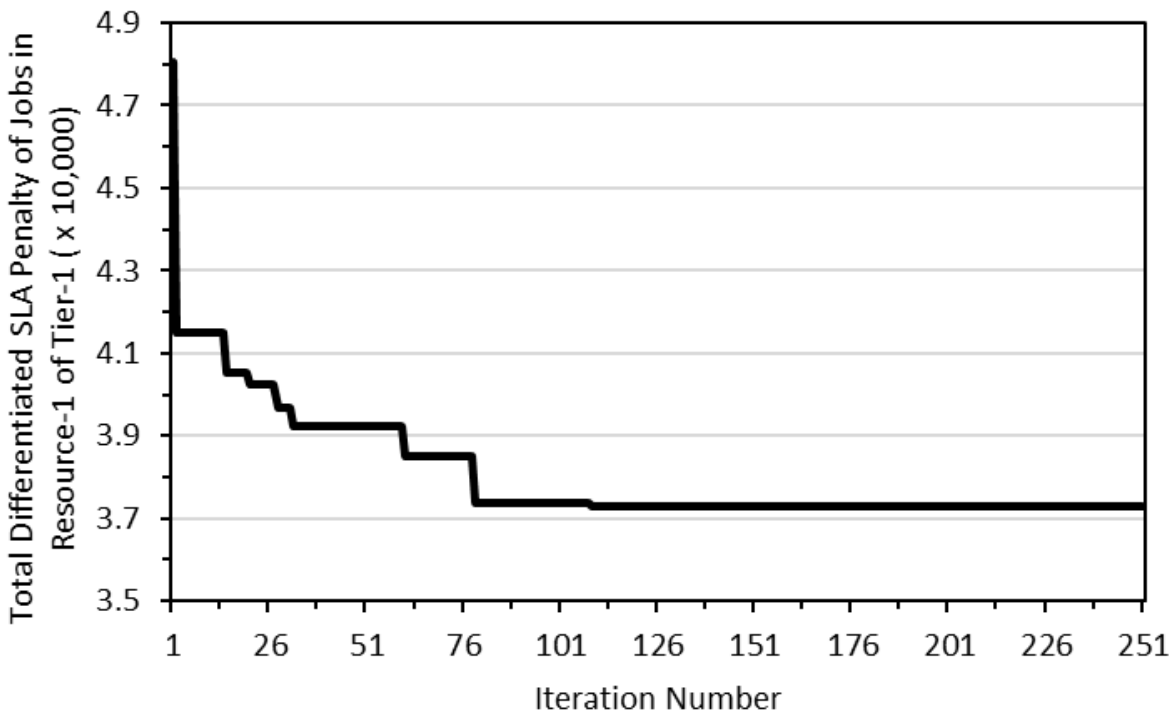}
	            \caption{Resource 1 of Tier 1\\~~~(Queue of 9 Jobs)}
	            \label{fig:Server1Tier1SYSTEMproportional_48049}
        \end{subfigure}
        ~
        \begin{subfigure}{0.315\textwidth}
		        \includegraphics[width=\textwidth]
                {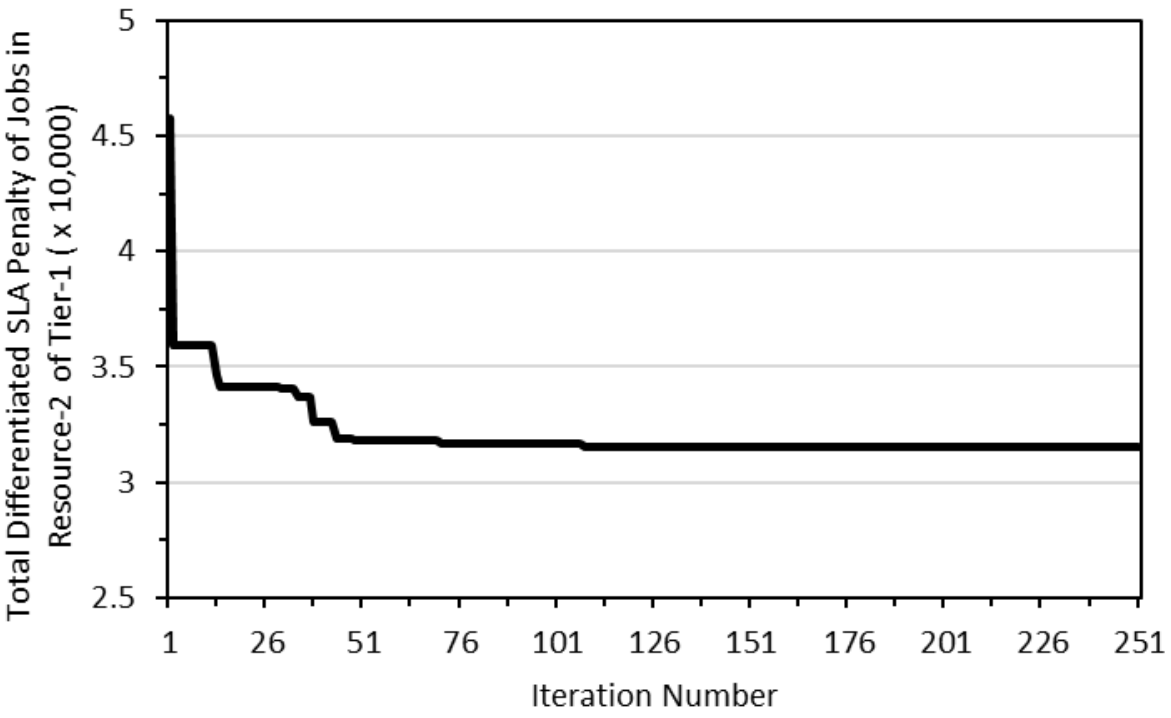}
	            \caption{Resource 2 of Tier 1\\~~~(Queue of 9 Jobs)}
	            \label{fig:Server2Tier1SYSTEMproportional_45752}
        \end{subfigure}
        ~
        \begin{subfigure}{0.315\textwidth}
                \includegraphics[width=\textwidth]
                {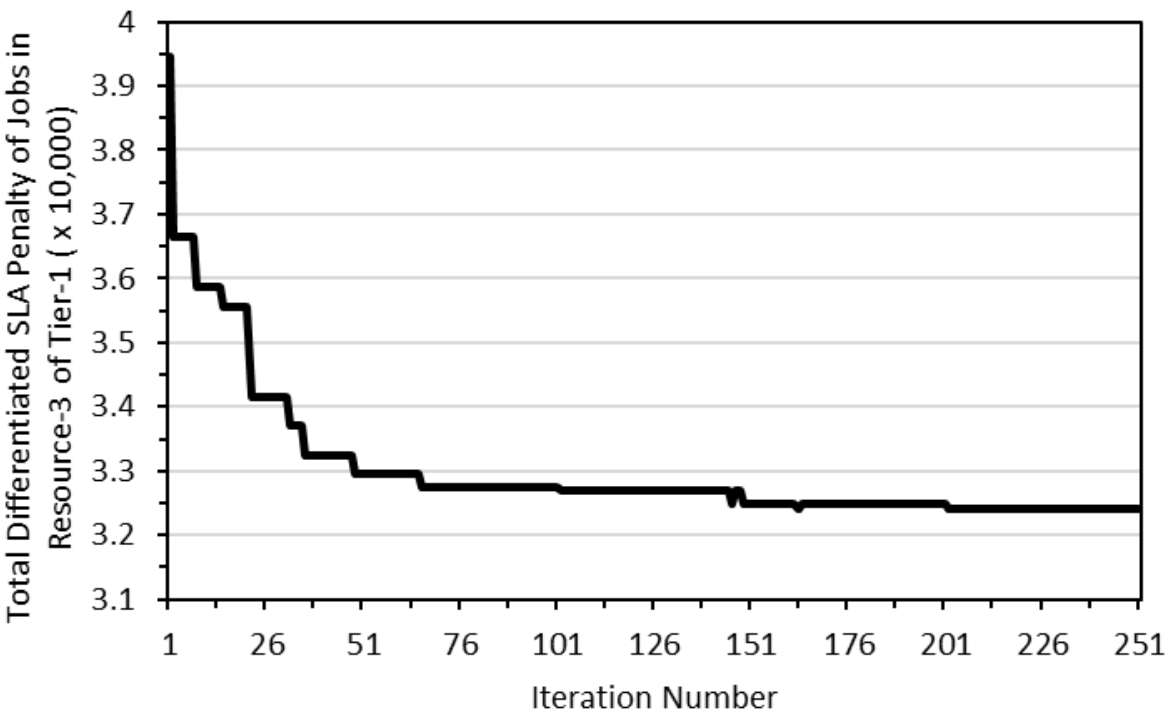}
	            \caption{Resource 3 of Tier 1\\~~~(Queue of 11 Jobs)}
	            \label{fig:Server3Tier1SYSTEMproportional_39447}
        \end{subfigure}

        \begin{subfigure}{0.315\textwidth}
		        \includegraphics[width=\textwidth]
                {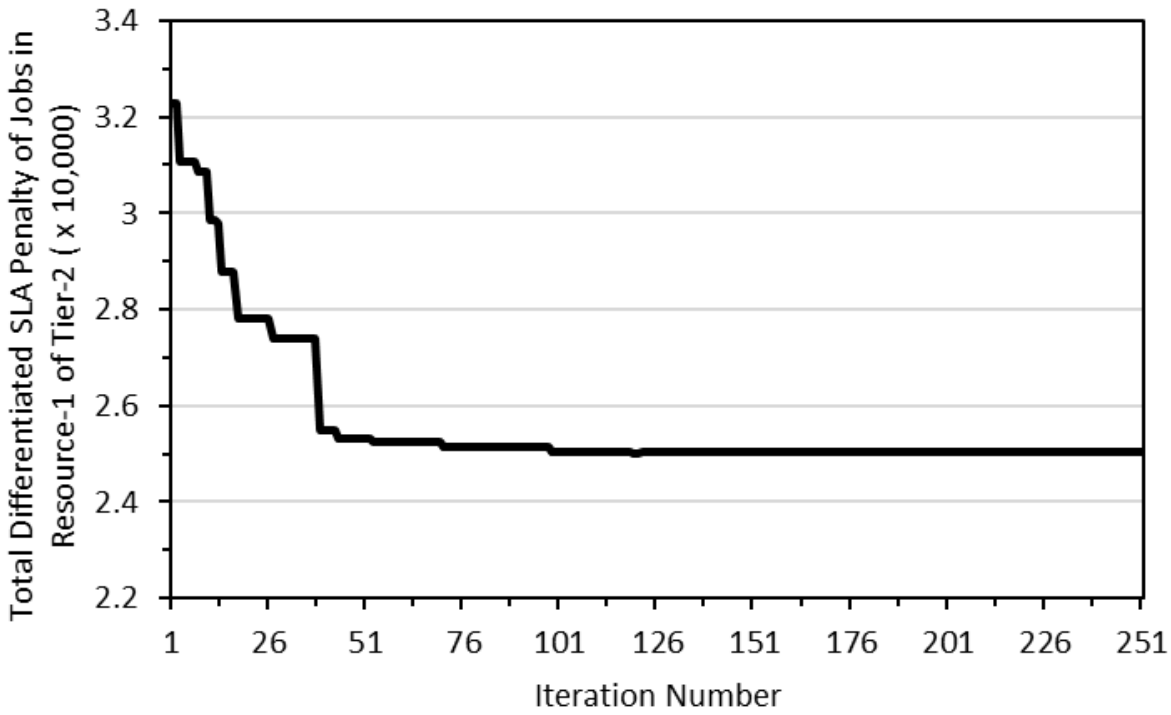}
	            \caption{Resource 1 of Tier 2\\~~~(Queue of 10 Jobs)}
	            \label{fig:Server1Tier2SYSTEMproportional_32291}
        \end{subfigure}
        ~
        \begin{subfigure}{0.315\textwidth}
		        \includegraphics[width=\textwidth]
                {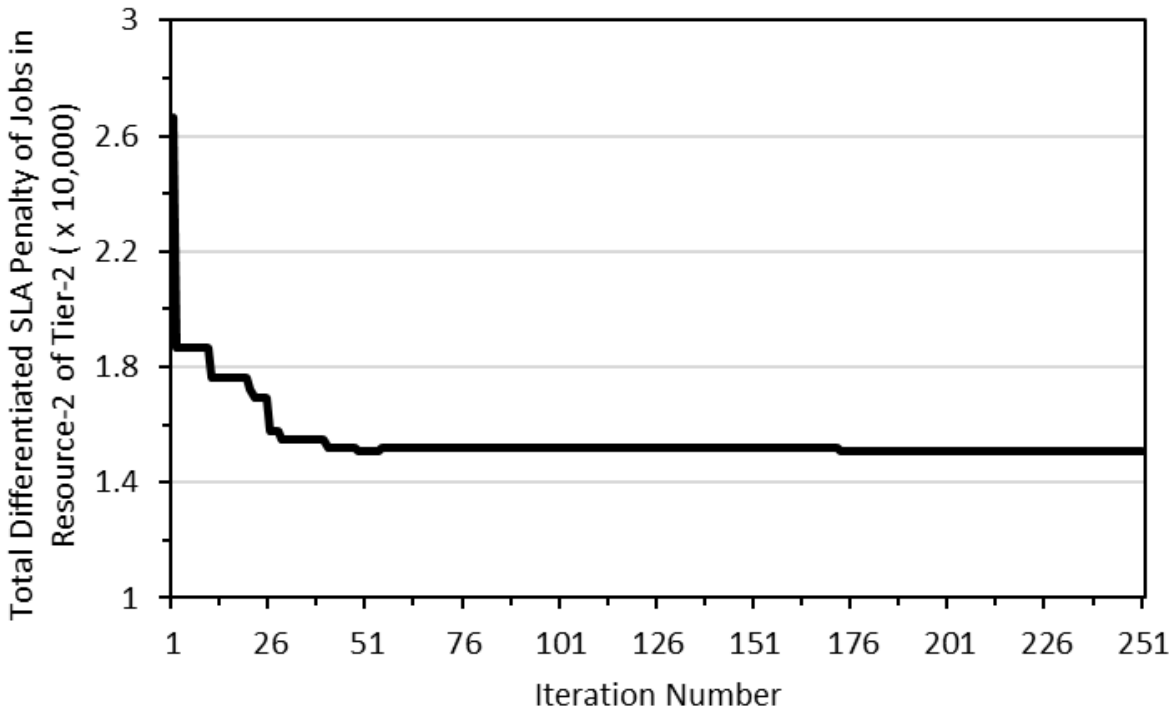}
	            \caption{Resource 2 of Tier 2\\~~~(Queue of 8 Jobs)}
	            \label{fig:Server2Tier2SYSTEMproportional_26629}
        \end{subfigure}
        ~
        \begin{subfigure}{0.315\textwidth}
                \includegraphics[width=\textwidth]
                {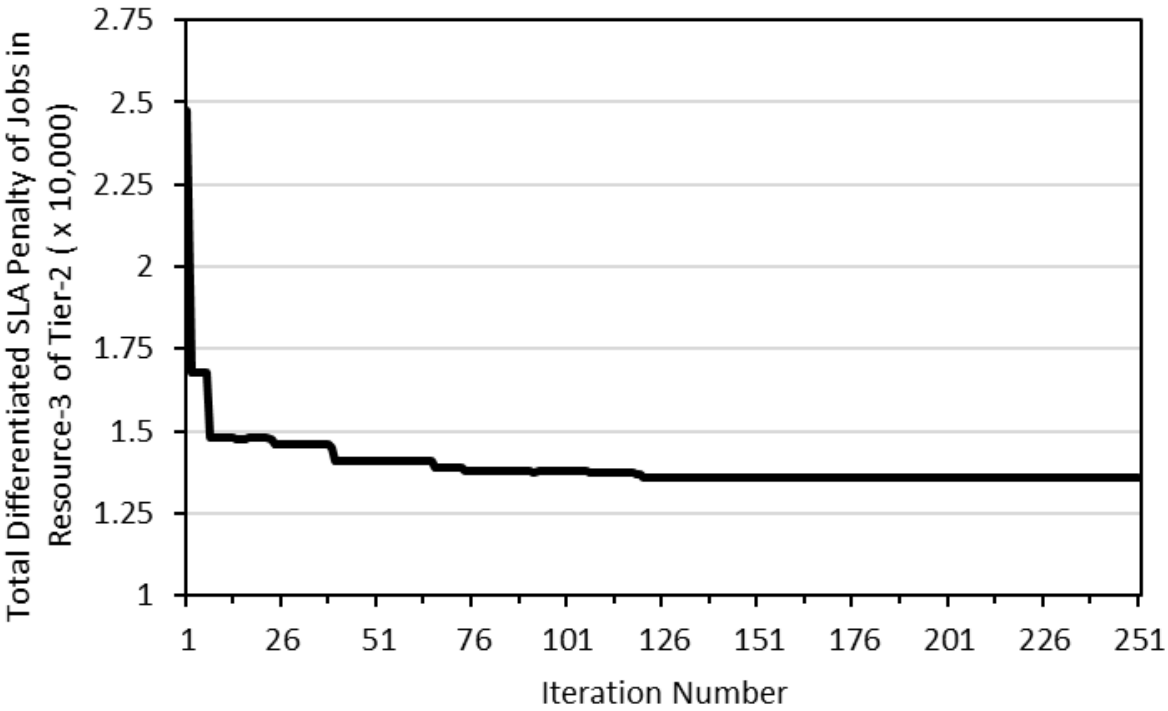}
	            \caption{Resource 3 of Tier 2\\~~~(Queue of 10 Jobs)}
	            \label{fig:Server3Tier2SYSTEMproportional_24726}
        \end{subfigure}

        \caption{Differentiated SLA Penalty $\omega\!\mathcal{P\!T}\!_{i,j}$ Based Segmented Queue Scheduling}
        \label{fig:Case4_DslaPQueue_SYSTEMdifferentiated}
\end{figure*}

\begin{table*}[!ht]
\label{tab:Queue1}
\caption{Differentiated SLA Penalty $\omega\!\mathcal{P\!T}\!_{i,j}$ Based Segmented Queue Scheduling}
\captionsetup{justification=centering}
\begin{center}\scalebox{0.8}{
\begin{threeparttable}
\begin{tabular}{c|c|cccc|cc}
\hline
\multirow{2}{*}{} & \multirow{2}{*}{\textbf{\begin{tabular}[c]{@{}c@{}}Number \\of Jobs\end{tabular}}}
& \multicolumn{2}{c}{\textbf{Initial}\tnote{1}} & \multicolumn{2}{c|}{\textbf{Enhanced}\tnote{2}}
& \multicolumn{2}{c}{\textbf{Improvement}}  \\ \cline{3-8}
& & \textbf{Violation}  & \textbf{Penalty} & \textbf{Violation} & \textbf{Penalty} & \textbf{Violation \%} & \textbf{Penalty \%} \\ \hline \hline

System-Level, Figure~\ref{fig:SYSTEMproportional_216897}
& 57 & 216897 & 1.80 & 154844 & 1.35 & 28.61\% & 25.35\%   \\ \hline

Resource-1 Tier-1, Figure~\ref{fig:Server1Tier1SYSTEMproportional_48049}
& 9  & 48050 & 0.38 & 37272 & 0.31 & 22.43\% & 18.45\%  \\
Resource-2 Tier-1, Figure~\ref{fig:Server2Tier1SYSTEMproportional_45752}
& 9  & 45753 & 0.37 & 31513 & 0.27 & 31.12\% & 26.38\%  \\
Resource-3 Tier-1, Figure~\ref{fig:Server3Tier1SYSTEMproportional_39447}
& 11 & 39447 & 0.33 & 32400 & 0.28 & 17.87\% & 15.10\%  \\ \hline

Resource-1 Tier-2, Figure~\ref{fig:Server1Tier2SYSTEMproportional_32291}
& 10 & 32291 & 0.28 & 24992 & 0.22 & 22.60\% & 19.87\%   \\
Resource-2 Tier-2, Figure~\ref{fig:Server2Tier2SYSTEMproportional_26629}
& 8  & 26630 & 0.23 & 15065 & 0.14 & 43.43\% & 40.18\%   \\
Resource-3 Tier-2, Figure~\ref{fig:Server3Tier2SYSTEMproportional_24726}
& 10 & 24726 & 0.22 & 13601 & 0.13 & 44.99\% & 41.95\%   \\ \hline

\end{tabular}
\begin{tablenotes}\scriptsize
\item[1] \textbf{Initial Violation} represents the total SLA violation time of jobs according to their initial scheduling before using the segmented queue genetic solution.
\item[2] \textbf{Enhanced Violation} represents the total SLA violation time of jobs according to their final/enhanced scheduling found after using the segmented queue genetic solution.
\end{tablenotes}
\end{threeparttable}}
\end{center}
\end{table*}

In the same way, the financial performance of the differentiated multi-tier penalty $\omega\!\mathcal{P\!T}\!_{i,j}$ based scheduling of the system virtualized queue genetic approach corroborates the financial performance of the former differentiated penalty $\omega\!\mathcal{A\!L}_i$ based scheduling. Cost-optimal schedules at the multi-tier level are also produced by the differentiated multi-tier penalty $\omega\!\mathcal{P\!T}\!_{i,j}$ based scheduling of the segmented queue genetic approach, which as well corroborates the financial performance of the differentiated multi-tier penalty $\omega\!\mathcal{A\!L}_i$ based scheduling of the segmented queue genetic approach.

For instance, the SLA penalty of the system-state shown in Table~6 and Figure~\ref{fig:Case4_TierDslaPSYSTEMproportional} is optimized at the multi-tier level by $22.05\%$, a reduction in the SLA penalty cost from $1.71$ for the initial schedule to reach $1.33$ for the improved schedule efficiently computed by the differentiated multi-tier penalty $\omega\!\mathcal{P\!T}\!_{i,j}$ based scheduling of the system virtualized queue genetic approach. In addition, the differentiated multi-tier penalty $\omega\!\mathcal{P\!T}\!_{i,j}$ based scheduling of the segmented queue genetic approach improves the financial performance of the SLA penalty by $25.35\%$ at the multi-tier level, which reduces the SLA penalty cost of the system-state from $1.8$ for the initial schedule to $1.35$ for the enhanced schedule shown in Table~7 and Figure~\ref{fig:Case4_DslaPQueue_SYSTEMdifferentiated}.

\begin{figure*}[!ht]
\centering
\captionsetup{justification=centering}
      \includegraphics[width=0.92\textwidth,height=0.2852\textheight]{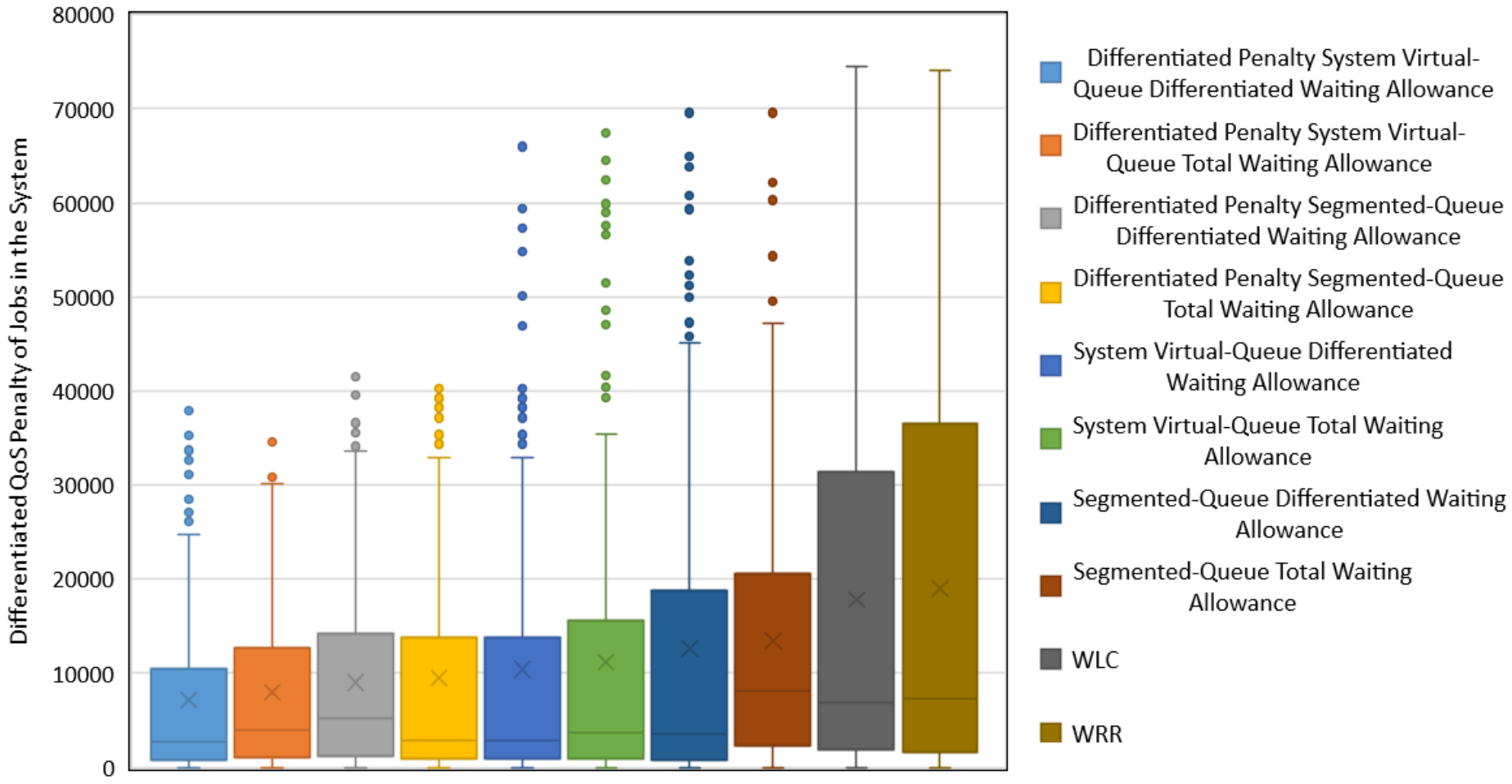}
	  \caption{Comparison of the Approaches}
      \label{fig:comparingThemTogether4}
\end{figure*}

A comparison of the financial performance of the differentiated penalty-based scheduling strategies in optimizing the differentiated SLA penalty cost at the multi-tier level is presented in Table~8 and Figure~\ref{fig:comparingThemTogether4}. The differentiated multi-tier penalty $\omega\!\mathcal{A\!L}_i$ based and $\omega\!\mathcal{P\!T}\!_{i,j}$ based scheduling efficiently produce optimal schedules that reduce the SLA penalty cost, using the system virtualized queue and segmented queue genetic scheduling solutions. However, compared with the differentiated service penalty scheduling approaches, the multi-tier $\omega\!\mathcal{A\!L}_i$ based and $\omega\!\mathcal{P\!T}\!_{i,j}$ based scheduling approaches demonstrate a superior performance in reducing the SLA penalty cost.

\begin{table*}[ht!]
\label{tab:comparison4}
\caption{Total Differentiated SLA Penalty}
\captionsetup{justification=centering}
\begin{center}\scalebox{0.6}{
\centering
\begin{tabular}{cc|cc|cc|cc|c|c}
\hline
\multicolumn{2}{c|}{\textbf{\begin{tabular}[c]{@{}c@{}}Differentiated Penalty Multi-Tier \\$\omega\!\mathcal{P\!T}\!\!_{i,j}$ Based Scheduling\end{tabular}}} &
\multicolumn{2}{c|}{\textbf{\begin{tabular}[c]{@{}c@{}}Differentiated Penalty Multi-Tier \\$\omega\!\mathcal{A\!L}_i$ Based Scheduling\end{tabular}}} &
\multicolumn{2}{c|}{\textbf{\begin{tabular}[c]{@{}c@{}}Multi-Tier \\$\omega\!\mathcal{P\!T}\!\!_{i,j}$ Based Scheduling\end{tabular}}} &
\multicolumn{2}{c|}{\textbf{\begin{tabular}[c]{@{}c@{}}Multi-Tier \\$\omega\!\mathcal{A\!L}_i$ Based Scheduling\end{tabular}}} &
\multirow{3}{*}{\textbf{WLC}} &
\multirow{3}{*}{\textbf{WRR}} \\ \cline{1-8}

\textbf{\begin{tabular}[c]{@{}c@{}}System \\Virtualized Queue\end{tabular}}        &
\textbf{\begin{tabular}[c]{@{}c@{}}Segmented \\Queue\end{tabular}}                 &
\textbf{\begin{tabular}[c]{@{}c@{}}System \\Virtualized Queue\end{tabular}}        &
\textbf{\begin{tabular}[c]{@{}c@{}}Segmented \\Queue\end{tabular}}                 &
\textbf{\begin{tabular}[c]{@{}c@{}}System \\Virtualized Queue\end{tabular}}        &
\textbf{\begin{tabular}[c]{@{}c@{}}Segmented \\Queue\end{tabular}}                 &
\textbf{\begin{tabular}[c]{@{}c@{}}System \\Virtualized Queue\end{tabular}}        &
\textbf{\begin{tabular}[c]{@{}c@{}}Segmented \\Queue\end{tabular}}                 &  &    \\ \hline

1431984 & 1800853 & 1589481 & 1897843 & 2074843 & 2521244 & 2228040 & 2692282 & 3559464 & 3805631   \\ \hline

\end{tabular}}
\end{center}
\end{table*} 

Differentiated multi-tier penalty $\omega\!\mathcal{A\!L}_i$ based scheduling of the segmented queue genetic approach reduces the SLA penalty by approximately $47\%$ compared with WLC and $50\%$ compared with WRR; however, it shows an inferior financial performance compared with the differentiated multi-tier penalty $\omega\!\mathcal{P\!T}\!_{i,j}$ based scheduling of the segmented queue genetic approach. In contrast, differentiated multi-tier penalty $\omega\!\mathcal{A\!L}_i$ based scheduling of the system virtualized queue genetic approach produces schedules that entail a cost of $\nolinebreak{1.59\!\times\!10^6}$ time units of the SLA penalty at the multi-tier level, a reduction of $55\%$ and $58\%$ compared with WLC and WRR strategies, respectively. Superior financial performance is demonstrated in the differentiated multi-tier penalty $\omega\!\mathcal{P\!T}\!_{i,j}$ based scheduling of the system virtualized queue genetic approach, which produces schedules that reduce the SLA penalty to around a cost of $\nolinebreak{1.43\!\times\!10^6}$ time units.

\section{Conclusion}
\label{sec:conc}

This paper presents a QoS-driven scheduling approach to address the differentiated penalty of delay-intolerant jobs in a multi-tier cloud computing environment. The approach emphasizes the notion of financial penalty in scheduling client jobs so that schedules are effectively produced based on economic considerations. Job treatment regimes are devised in a differentiated QoS penalty model, so as the cloud service provider computes schedules that capture the financial impact of SLA violation penalty on the QoS provided. Optimal performance is delivered to clients who cannot afford the cost of SLA violations and delays.

The proposed queue virtualization design schemes facilitate the formulation of schedules at the tier and multi-tier levels of the cloud computing environment. The design schemes leverage the utilization of resources within a tier to derive tier-driven schedules with optimal performance, as well as employ dependencies and bottleneck shifting between tiers to formulate multi-tier-driven schedules with optimal performance. The proposed meta-heuristic approaches, represented by the differentiated penalty virtualized-queue and segmented-queue genetic solutions, reduce the complexity of optimal scheduling of jobs on resource queues of the tiers.

The formulated cost-optimal schedules reduce the cost of SLA penalty for client jobs, which accordingly maximizes client satisfactions and thus loyalties to the cloud service provider. The produced schedules maintain a balance between delivering the highest QoS provided to clients while ensuring an efficient system performance with a reduced operational cost, and thus fulfilling the different QoS expectations and mitigating their associated commercial penalties. It is shown that the financial performance has been improved by reducing the QoS penalty under different SLA commitments of client jobs in a multi-tier cloud computing environment.

\section{Future Work}
\label{sec:future}

A cloud service provider employs multiple resources that typically demand a huge amount of energy to execute various client demands. Due to its impact on system performance, energy saving has recently become of paramount importance in cloud computing. However, a major challenge on a cloud service provider is maintaining a maximum energy efficiency (minimum consumption) while ensuring high system performance that fulfills the different QoS expectations in executing client jobs of varying computational demands. Any imbalance in managing these conflicting objectives may result in failing to meet SLA obligations of clients and, thus, financial penalties on the cloud service provider. Accordingly, it is imperative to devise scheduling approaches that produce energy-efficient optimal schedules with minimal SLA penalties at the multi-tier level. A sustainable cloud computing environment would help reduce the energy cost required to execute client demands.

\bibliographystyle{IEEEtran}

\bibliography{\jobname}
\end{document}